\documentclass[fleqn,usenatbib]{mnras}

\usepackage{newtxtext,newtxmath, caption, subfig}
 
\usepackage{appendix}
\usepackage{float}

\usepackage[T1]{fontenc}

\DeclareRobustCommand{\VAN}[3]{#2}
\let\VANthebibliography\thebibliography
\def\thebibliography{\DeclareRobustCommand{\VAN}[3]{##3}\VANthebibliography}

\usepackage{graphicx}	
\usepackage{amsmath}	
\usepackage{amssymb}	

\newcommand{\lya}{\mathrm{{Ly\alpha}}}

\title[Late Reionization and LAEs]{Constraining Reionization in Progress at $z=5.7$ with Lyman-$\alpha$ Emitters: Voids, Peaks, and Cosmic Variance}

\author[Nakul Gangolli]{
Nakul Gangolli$^{1}$\thanks{E-mail: ngang002@ucr.edu},
Anson D'Aloisio$^{1}$\thanks{E-mail: ansond@ucr.edu},
Fahad Nasir$^{1}$\thanks{E-mail: fahadn@ucr.edu}, \&
Zheng Zheng$^{2}$\thanks{E-mail: zhengzheng@astro.utah.edu} \smallskip \\
$^{1}$Department of Physics and Astronomy, University of California, Riverside, CA 92521, USA \\
$^{2}$Department of Physics and Astronomy, University of Utah, Salt Lake City, UT 84112, USA
}

\pubyear{2020}

\begin{document}
\label{firstpage}
\pagerange{\pageref{firstpage}--\pageref{lastpage}}
\maketitle

\begin{abstract}
A number of independent observations suggest that the intergalactic medium was significantly neutral at $z=7$ and that reionization was, perhaps, still in progress at $z=5.7$. The narrowband survey, SILVERRUSH, has mapped over $2,000$ Lyman-$\alpha$ emitters (LAEs) at these redshifts.  Previous analyses have assumed that reionization was over by $z=5.7$, but this data may actually sample the final stages of reionization when the last neutral islands were relegated to the cosmic voids. Motivated by these developments, we reexamine LAE void and peak statistics and their ability to constrain reionization.  We construct models of the LAE distribution in (1 Gpc$/h$)$^3$ volumes, spanning a range of neutral fractions at $z=5.7$ and 6.6.  Models with a higher neutral fraction show an enhanced probability of finding holes in the LAE distribution.  When comparing models at fixed mean surface density, however, LAEs obscured by neutral gas in the voids must be compensated by visible LAEs elsewhere. Hence, in these models the likelihood of finding an over-dense peak is also enhanced in the latter half of reionization. Compared to the widely used angular two-point correlation function (2PCF), we find that the void probability function (VPF) provides a more sensitive test of models during the latter half of reionization.  By comparison, at neutral fractions $\sim 50\%$, the VPF and a simple peak thresholding statistic are both similar to the 2PCF in constraining power.  Lastly, we find that the cosmic variance and large-scale asymmetries observed in the SILVERRUSH fields are consistent with large-scale structure in a $\Lambda$CDM universe.  
\end{abstract}




\section{Introduction}

Observational measurements have narrowed the range of viable reionization models to those that end around $z=6$.  The Planck measurement of the Thomson scattering optical depth to the Cosmic Microwave Background (CMB) places the bulk of reionization between $z=6-12$ \citep{2020A&A...641A...6P}.  This timing is consistent with transmission measurements of the Ly$\alpha$ and Ly$\beta$ forests \citep{fan06, mcgreer15,2019ApJ...884...30W}, as well as evidence of quasar damping wings in $z>7$ quasars  \citep{mortlock11, 2018ApJ...864..142D, 2020ApJ...896...23W}.  Additionally, at $z\gtrsim 5.5$, the Ly$\alpha$ forest exhibits large scatter in its transmission averaged over long path lengths, epitomized by the $110 h^{-1}$ comoving Mpc Ly$\alpha$ trough reported by \citealt{2015MNRAS.447.3402B} (see also \citealt{fan06,2018MNRAS.479.1055B, 2018ApJ...864...53E}). Some models proposed to explain this scatter place the end of reionization as late as $z=5$ \citep{2019MNRAS.485L..24K, 2019arXiv191003570N, 2020MNRAS.491.1736K}.  Although a broadly consistent picture of late reionization has emerged from all of these observations, the reionization history in detail as well as its sources remain unknown. 

LAEs have long been an important probe of the EoR owing to the resonant nature of the $\lya$ line to neutral gas.   Visible LAEs indicate the presence of ionized gas within the IGM because even small densities of neutral hydrogen nearby can attenuate LAEs significantly. Indeed, the luminosity function of LAEs evolves rapidly at $z>6$ and the fraction of Lyman Break Galaxies showing strong Ly$\alpha$ emission is also observed to decline \citep[e.g.][]{2006ApJ...648....7K, schenker12, ono12, 2014ApJ...793..113P, 2015MNRAS.446..566M, 2016ApJ...818...38S, 2018ApJ...856....2M, Inoue_2018, 2019ApJ...886...90H}.  The drop-off in space density of LAEs can be attributed to an increasingly neutral IGM toward high redshift, an interpretation bolstered by other observations as described above \citep[see however][]{2017ApJ...839...44S, 2020arXiv201000023H}. The differential evolution of the bright and faint LAE populations at $z>7$ \citep[e.g.][]{2017MNRAS.464..469S, 2020arXiv201003566E} appears consistent with expectations from reionization simulations.  In fact, models suggest that the brightest LAEs at $z>7$ may serve as signposts for the earliest ionized structures \citep{2018MNRAS.479.2564W, 2018ApJ...857L..11M}. A noteworthy case is the double-peaked emitter COLA1, for which the presence of a prominent blue wing places a lower limit on the size of the ionized bubble hosting it \citep{2016ApJ...825L...7H, 2018A&A...619A.136M, 2020MNRAS.tmp.2734M}.  LAEs have also been used in combination with Ly$\alpha$ forest data to study the connection between LAEs and their intergalactic environments \citep{2018ApJ...863...92B, 2018MNRAS.479...43K, 2020MNRAS.494.1560M}.

Given sufficiently large survey area, the angular two point correlation function (2PCF) is thought to be the most robust probe of reionization amongst LAE statistics.  Simulations show that the obscuration of LAEs by neutral gas during reionization generally enhances their apparent clustering \citep{2006MNRAS.365.1012F, 2007MNRAS.381...75M, 2014MNRAS.444.2114J, 2015MNRAS.450.4025H, 2015arXiv150502787S}. An early measurement of the 2PCF at $z=6.6$ from $207$ LAEs in $1$ deg$^2$ constrained their bias, host halo masses, and duty cycle \citep{2010ApJ...723..869O}.  The recently conducted narrowband survey, SILVERRUSH, greatly expanded the sample of known reionization-era LAEs using the Subaru/Hyper Suprime-Cam (HSC). The collaboration has reported over $2,000$ LAEs at $z=5.7$ and $6.6$ covering areas of $14-21$ deg$^2$ ($0.3-0.5$ Gpc$^2$).  With this updated sample, \citet{2018PASJ...70S..13O} found only mild redshift evolution in the 2PCF from $z=5.7$ to 6.6.  Comparing their measurements to previously published models, they estimated a global neutral fraction of $x_{\rm HI} = 0.15\pm 0.15$ at $z=6.6$.    At face value, this measurement provides little evidence for the late reionization scenario that has become increasingly supported by independent observations, although the uncertainties in the 2PCF remain large \citep[see e.g. discussion in ][]{Inoue_2018}. 

One of the central questions motivating the current paper is whether the 2PCF is the most sensitive statistic for constraining reionization during its final stages. As noted above, there is mounting evidence that existing LAE samples at $z\sim 7$ are probing deep into reionization.  Even the most densely sampled data at $z=5.7$ may be probing the tail end of reionization \citep{2019arXiv191003570N, 2020MNRAS.491.1736K}.  Radiative transfer simulations indicate that neutral gas becomes increasingly relegated to the voids in the source distribution when the global neutral fraction drops below $50\%$ \citep[e.g.][]{2019ApJ...870...18D, 2019MNRAS.485L..24K, 2019MNRAS.489.1590G, 2019MNRAS.490.3177W}. We therefore expect LAEs located in voids to be preferentially obscured by neutral islands during the last stages of reionization.  Motivated by this expectation, we reexamine the use of LAE void statistics for constraining reionization, with an eye towards recent models that place the neutral fraction at $\sim 10\%$ at $z=5.7$.    We also explore peaks in the LAE distribution as a complementary method for quantifying reionization's effect on the apparent clustering of LAEs.      

LAE void statistics in the context of reionization were considered by \citet{2007MNRAS.381...75M}.  The current paper adds to this work in three ways: (1) Whereas the simulations in \citet{2007MNRAS.381...75M} concluded reionization well before $z=5.7$, we focus explicitly on the possibility of constraining models in which reionization was still ongoing at $z=5.7$. This updates previous work with the most recent empirical developments and may be of particular interest because LAE samples are currently largest at $z=5.7$; (2) We employ simulation volumes of (1 Gpc$/h$)$^3$, a factor of $\approx 470$ bigger than the largest in \citet{2007MNRAS.381...75M},  to model reionization and the LAE population.  This allows us to develop statistically representative models of narrow band surveys whose individual fields of view often are comparable to, or larger than, the simulation box sizes of previous studies.  The SILVERRUSH fields are striking visually because they exhibit significant field-to-field variation in surface density, as well as clear asymmetries in the LAE distribution over enormous scales ($\sim 200$ Mpc$/h$). Some fields are pocketed with highly-clustered regions and nearly empty holes. We exploit our large simulation volumes to address additionally whether the observed features are consistent with expected cosmic variance from large-scale structure, or whether they may be indicative of some other source of variance; (3) We contrast the constraining power of voids and peak statistics with that of the 2PCF. We will find that voids provide a more sensitive test of models than the 2PCF during the late stages of reionization.   

The remainder of this paper is structured as follows.  In \S \ref{sec:method} we describe our simulation approach and in \S \ref{sec:models} we lay out and calibrate our late reionization models.  Section \ref{sec:stats} contains our main results on void and peak statistics, and cosmic variance.  In \S \ref{sec:conc} we offer concluding remarks.  We assume a standard $\Lambda$CDM cosmology with $H_0 = 100 h$ km/s/Mpc and $h=0.68$, $\Omega_m = 0.3$, and $\Omega_{\Lambda} = 0.7$, consistent with the latest Planck measurements \citep{2020A&A...641A...6P}. Unless otherwise stated, all distances are reported in comoving units and velocities are in proper units.

\section{Simulation Methodology}
\label{sec:method}

The visibility of LAEs is set by physical processes spanning an enormous dynamic range, from hundreds of Mpc$/h$ to sub-galactic scales. In addition, narrow band survey fields of view span ($\sim $hundreds of Mpc)$^2$. Our multi-scale approach attempts to bridge the scale gap by using several cosmological simulations in tandem.  In this section we describe each component of our models.

\subsection{The underlying LAE population}
\label{sec:LAE_pop}

The basis for our mock LAE catalogs is the publicly available halo catalogs from the Multi-Dark Simulation (MDPL) of \citep{2016MNRAS.457.4340K}, which affords us a 1 $(h^{-1}$ Gpc)$^3$ volume. The MDPL simulation is a dark-matter-only N-body simulation that was run with a modified version of the GADGET-2 code \citep{2005MNRAS.364.1105S} using $N=3840^3$ particles, which corresponds to a particle mass of $1.51\times 10^9h^{-1}$ M$_{\odot}$.  We used the friends-of-friends (FoF) halo catalogs provided by the \url{CosmoSim} database\footnote{\url{https://www.cosmosim.org/}}. The minimum halo mass in these catalogs is $M_{\rm min} = 3\times 10^{10} h^{-1}$ M$_{\odot}$.

To populate the MDPL halos with galaxies, we applied an abundance matching scheme that equates the cumulative number density of halos with the integrated rest-frame UV luminosity functions of \citet{2015ApJ...803...34B} (see also \citealt{2015ApJ...810...71F}). Following \citet{2019MNRAS.485.1350W}, we model the duty cycle of these galaxies with the method of \citet{2010ApJ...714L.202T}. The duty cycle is parameterized by a star formation time-scale $\Delta t$, which we take to be one of the free parameters in our models.  Smaller values of $\Delta t$ result in UV luminous galaxies being assigned to less massive halos. In what follows, we adopt $\Delta t = 50$ Myr as our fiducial value. In \S \ref{sec:models} we show that this choice provides a reasonable match to statistics of the observed LAE population.        

Given UV luminosities, the Ly$\alpha$ rest-frame equivalent widths (REWs) of galaxies were drawn randomly from the empirically calibrated model distribution of \citet{2012MNRAS.419.3181D}. Scatterings of Ly$\alpha$ photons by the interstellar gas results in a complex frequency structure for the Ly$\alpha$ line as it emerges from source galaxies.  We do not attempt to model these physical processes in detail here.  Instead we assume that the structure of the line blue-ward of systemic is completely absorbed by the IGM -- a good approximation at the high redshifts of interest, in most cases \citep[see however][]{2016ApJ...825L...7H, 2018A&A...619A.136M, 2018ApJ...859...91S}.  We model the red-ward side with a Gaussian profile characterized by a velocity offset, $\Delta v$, and width $\sigma_v$.  We set $\Delta v = \beta v_{\rm circ}$, where $v_{\rm circ}$ is the circular velocity of the halo at its boundary\footnote{Here, the ``radius" of an FoF halo is defined to be that of a sphere with equal volume.}, and $\beta$ is a free parameter in our model \citep[e.g.][]{2019MNRAS.485.1350W}.  Following \citet{2018ApJ...856....2M}, we fix $\sigma_v$ by setting the full-width half-maximum of the Gaussian equal to $v_{\rm circ}$.  These choices are motivated by radiative transfer calculations of the Ly$\alpha$ line structure emerging from simple models of the ISM in star-forming galaxies \citep{2002ApJ...578...33Z}.

\subsection{Opacity of the IGM}

The attenuation of Ly$\alpha$ emission lines by intergalactic hydrogen comes from two sources: (1) The neutral hydrogen in yet-to-be reionized gas; (2) Self-shielding regions and residual neutral hydrogen within ionized bubbles.  Sections \ref{sec:ESMR} and \ref{sec:SSRs}, respectively, describe how we model these. 

\subsubsection{Reionization simulations}
\label{sec:ESMR}

We modeled the distribution of neutral hydrogen during reionization with a numerical implementation of the excursion-set model of reionization (ESMR; \citealt{2004ApJ...613....1F, 2011MNRAS.411..955M}). Following previous implementations \citep[see e.g.][]{2011MNRAS.414..727Z}, we smoothed the MDPL halo field onto a grid with dimension $N = 512^3$ to obtain the collapsed fraction field, $f_{\rm coll} (\bf{r})$.  The latter was then smoothed over a hierarchy of scales with a top-hat filter in Fourier space.  Cells were marked as ionized when the condition $\zeta f_{\rm coll}({\bf r}) \geq 1$ was satisfied, where $\zeta$ is a free parameter quantifying the efficiency at which the sources deliver ionizing photons to the IGM.  As we describe in \S \ref{sec:models}, we adjusted $\zeta$ to obtain different global ionized fractions at $z=5.7$ and $z=6.6$.

It is important to account for the enhancement of the ionizing ultraviolet background (UVB) intensity around bright and/or clustered LAEs.  We modelled spatial fluctuations in the UVB with a simple attenuation model, separately from the ESMR reionization simulations.  We computed the hydrogen photoionization rate, $\Gamma_{\rm HI}$, on a uniform grid with $N=512^3$ cells assuming that each source at location $\vec{x}_j$ makes a contribution $\propto L_{912} \exp(-|\vec{x}_i - \vec{x}_j|/\lambda^{\rm mfp}_{912})/|\vec{x}_i - \vec{x}_j|^2$ to the cell at $\vec{x}_i$, where $L_{912}$ is the source's specific luminosity at 912 \AA, and $\lambda^{\rm mfp}_{912}$ is the mean free path.  For simplicity we adopt a uniform $\lambda^{\rm mfp}_{912} = 30 (15) h^{-1}$ Mpc at $z=5.7 (6.6)$, which is motivated by extrapolating the observational measurements of \citet{2014MNRAS.445.1745W} at $z \lesssim 5.2$.  Assuming a constant $\lambda^{\rm mfp}_{912}$ affords us the ability to compute $\Gamma_{\rm HI}$ fields efficiently using Fast Fourier Transforms, at the expense of neglecting potentially large spatial variations in $\lambda^{\rm mfp}_{912}$ at these redshifts \citep{2016MNRAS.460.1328D, 2018MNRAS.473..560D}.  Small local values of $\lambda^{\rm mfp}_{912}$ -- for example, in cosmic voids with a dearth of sources -- leads to more attenuation of the local LAE population by self-shielding systems \citep{2013MNRAS.429.1695B, 2017ApJ...839...44S}, introducing another source of fluctuation in the LAE distribution.  Another effect that our method misses is the shadowing of the UVB which leads to a suppression of $\Gamma_{\rm HI}$ in the vicinity of neutral hydrogen \citep{2019arXiv191003570N}.       

\subsubsection{Absorptions within ionized bubbles }
\label{sec:SSRs}

Even within an ionized bubble, residual neutral hydrogen in the cosmic web can contribute significantly to the attenuation of Ly$\alpha$ emission \citep[e.g.][]{2013MNRAS.429.1695B, 2015MNRAS.446..566M}.  We model this component of the opacity using one-dimensional skewers extracted from an Eulerian cosmological hydrodynamics simulation run with a modified version of the code of \citet{2004NewA....9..443T}. The details of our code implementation have been described elsewhere \citep{2018MNRAS.473..560D, 2019arXiv191003570N}. Here we summarize the salient points (including updates) for the current paper.  

Our simulation used a periodic box with side length $L_{\rm box} = 20 h^{-1}$ Mpc, and $2\times 2048^3$ gas and dark matter resolution elements.  The simulation was initiailzed at $z=300$ using first-order perturbation theory and transfer functions from the CAMB software \citep{Lewis:1999bs}.  The gas was flash ionized at redshift $z_{\rm reion} = 8.5$ by a uniform ionizing radiation background with specific intensity $J_{\nu} \propto \nu^{-1.5}$ (where $\nu$ is frequency) between $1-4$ Ry. At this time the gas was also impulsively heated to $T = 20,000$ K.  The normalization of the UVB was fixed to give a global mean hydrogen photoionization rate of $\Gamma_{\rm HI} = 10^{-13}$ s$^{-1}$.  However, to incorporate spatial fluctuations in the UVB, we re-scaled $\Gamma_{\rm HI}$ in post-processing along the skewers using the UVB models from the last section. To model self-shielding in over-dense regions, we implemented the model of \citet{2013MNRAS.430.2427R} with parameters updated by \citet{2018MNRAS.478.1065C}.  We traced 20,000 skewers at random angles from halos\footnote{Halos were identified with a spherical-overdensity criterion in which the mass enclosed is $M_{200} = 200 \bar{\rho}_m (4 \pi /3) R^3_{200}$, where $\bar{\rho}_m$ is the cosmic mean matter density and $R_{200}$ is the radius within which the mean matter density is $ 200 \bar{\rho}_m$.} with masses $M\geq 10^{10}$ M$_{\odot}$. As we describe in the next section, we used the gas data from these skewers to model attenuation of the Ly$\alpha$ line by self-shielding regions along the sight lines to LAEs.

\subsection{Construction of Mock Surveys}
\label{sec:mock_surveys}

We piece together all the elements described above to construct mock LAE catalogs.  The first step is to create $1,000$ mock fields of view (FoVs) by taking randomly located and oriented sub-volumes of the MDPL simulation -- utilizing the periodic boundary conditions. Our sub-volumes have widths of 180(200) $h^{-1}$Mpc at $z=5.7(6.6)$, and depths of $29~h^{-1}$ Mpc.   The widths correspond approximately to the average FoV scale of SILVERRUSH, and the depths represent the selection windows of the NB816 (NB921) filters \citep{2018PASJ...70S..13O}.  We then populate the halos in these sub-volumes with LAEs using the method outlined in \S \ref{sec:LAE_pop}.  This step produces the ``intrinsic" LAE population.   

The next step is to attenuate Ly$\alpha$ luminosities and REWs by applying the opacity of the IGM redward of systematic $\lya$.  We trace skewers of length $150~h^{-1}$Mpc from each LAE along the line of sight.  The skewers consist of uniform grid points at the resolution of the hydrodynamics simulation, $\Delta x = 9.77~h^{-1}$kpc, or $\Delta v = 1.41 (1.50) $ km/s at $z=5.7(6.6)$. Gas properties are assigned to these skewers from randomly chosen skewers through the hydrodynamics simulation. Our ESMR simulations provide the morphology of the neutral gas around the LAEs.  For a given reionization model, wherever a skewer intersected neutral regions in the corresponding ESMR field, the neutral fraction of the gas was set to unity. We emphasize that the ESMR fields and the LAEs share the same underlying density field (from the MDPL box), but the $L=20~h^{-1}$Mpc hydrodynamics simulation does not.  Similar multi-scale approaches have been employed by \citet{2015MNRAS.446..566M} and \citet{2018ApJ...864..142D}.      

By integrating along the line of sight, we compute the IGM transmission fraction redward of systemic, 
\begin{equation}
    T_{\lya}^\mathrm{IGM}  = \frac{\int d\nu J(\nu) e^{-\tau(\nu)}}{\int d\nu J(\nu) }, 
\end{equation}
where $J(\nu)$ is the line profile normalized to intergrate to unity, and $\tau({\nu})$ is the opacity including contributions from the ionized and fully neutral regions \citep{2015MNRAS.446..566M, 2019MNRAS.485.1350W}.  After attenuating the LAEs, we apply cuts on the Ly$\alpha$ luminosity ($L_{\mathrm{Ly}\alpha}$) and REW.  Unless specified otherwise, we select $L_{\mathrm{Ly}\alpha} > 6.3\times 10^{42}$ ($7.9\times 10^{42}$) ergs/s and REW $> 10 (14)$ \AA\ at $z=5.7 (6.6)$.  These cuts from \citet{2019MNRAS.485.1350W} approximate those employed by SILVERRUSH, though we note that different sample cuts are applied in the SILVERRUSH analyses depending on the application.  We also note that different SILVERRUSH fields have different limiting depths. 

The end result of this process is a sample of $1,000$ mock survey FoVs with properties similar to the SILVERRUSH fields. We will use these mocks to explore the effect of reionization on various LAE statistics in \S \ref{sec:stats}.

\begin{table}
    \centering
    \begin{tabular}{|c|c|c|c|c|c|c}
         z & $\bar{x}_{\rm HI}$ & $\beta$ & $\lambda_{912 \AA}^{\mathrm{mfp}}$  & $\langle \Gamma_{\mathrm{HI}}\rangle$ & $\Delta t$  & $\langle \Sigma \rangle$ \\
          & & & [cMpc/$h$] & [$\times10^{-12}$ s$^{-1}$] & [Myr] &  [deg$^{-2}$] \\
         \hline \hline
         
         5.7& 0.0 & 1.2  & 30 & 0.37 & 50 & 72 \\ \hline
         & 0.1 & 1.25 & 30 & 0.50 & 50 & 72 \\ \hline
         & 0.3 & 1.4  & 30 & 0.74 & 50 & 71 \\ \hline \hline

         6.6 & 0.0 & 1.2  & 15 & 0.37 & 50 & 32 \\ \hline
         & 0.25 & 1.4 & 15 & 0.50 & 50 & 32 \\ \hline
         & 0.5 & 1.8  & 15 & 0.74 & 50 & 32 \\ \hline
    \end{tabular}
    \caption{Summary of our model parameters.  Columns 2-6 list the five free parameters in our models. The last column shows the resulting mean LAE surface densities, $\langle \Sigma \rangle$, which have been tuned to match (approximately) the observed values in SILVERRUSH.  Here, $\bar{x}_{\rm HI}$ is the mean neutral fraction of the IGM, $\beta$ is the proportionality constant between halo circular velocity and Ly$\alpha$ line offset, $\lambda_{912 \AA}^{\mathrm{mfp}}$ is the mean free path in ionized regions, $\langle \Gamma_{\rm HI} \rangle$ is the mean hydrogen photoionization rate in ionized regions, and $\Delta t$ is the duty cycle star formation timescale. }
    \label{tab:beta_coeff}
\end{table}

\section{Models}
\label{sec:models}

\begin{figure*}
    \centering
    \includegraphics[width=\textwidth]{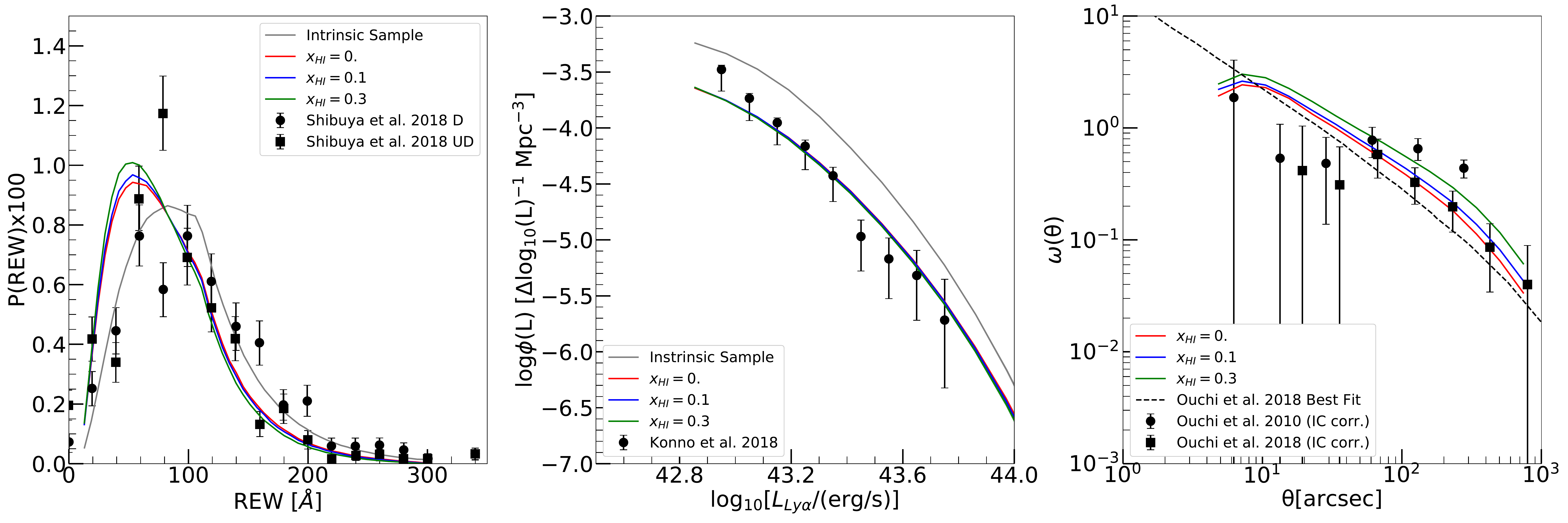}
    \includegraphics[width=\textwidth]{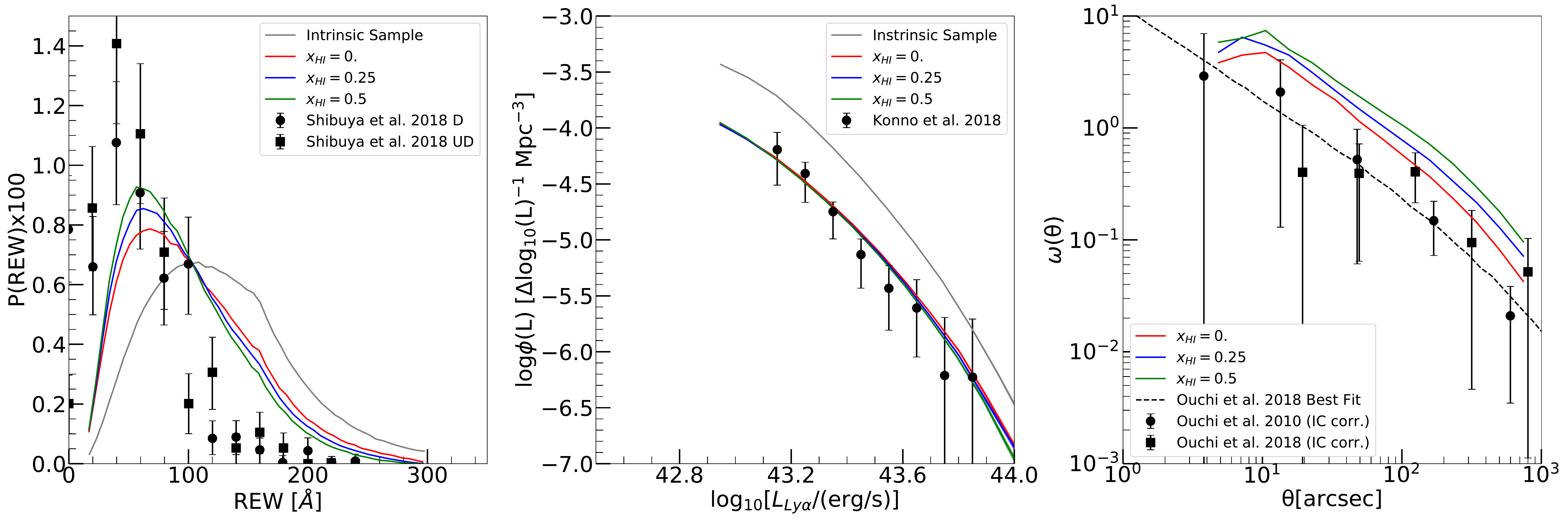}
    \captionsetup{width=\textwidth}
    \caption{Comparison of our models to recent observational measurements.  The left, middle, and right columns correspond to the REW distribution, Ly$\alpha$ luminosity function, and angular two-point correlation function, respectively.  The top and bottom rows show results for $z=5.7$ and $6.6$, respectively. Our models are distinguished by the mean neutral fraction of the IGM, as indicated in the plot legends.  In the left and middle panels, the light gray curves correspond to the ``intrinsic" LAE population, before any attenuation from the IGM/neutral regions is applied.  In the right panels, the dashed curves show the best fit power-law models of \citet{2018PASJ...70S..13O}.  In the left panel legends, ``D'' and ``UD'' denote the Deep and the Ultra-Deep fields, respectively.  Our models parameters have been tuned to provide reasonable agreement with these observations.  In \S \ref{sec:stats}, we use these models to explore the efficacy of void and peak statistics for constraining reionization.  }
    \label{fig:obs_anchors_57}
\end{figure*}

A summary of parameters for each of our models is provided in Table \ref{tab:beta_coeff}. We generated mock LAE populations for three global neutral fractions ($\bar{x}_{\mathrm{HI}}$) at $z=5.7$, and three at $z=6.6$. Since the neutral fraction is the key distinction between our models, we will label them with $\bar{x}_{\mathrm{HI}}$ hereafter.  The values of $\bar{x}_{\mathrm{HI}}$ are loosely motivated by the ``late" and ``early" reionization scenarios considered in \citet{2019arXiv191003570N}. We note that a neutral fraction of $\bar{x}_{\mathrm{HI}}=0.3$ at $z=5.7$ is in mild tension with the Ly$\alpha$ forest dark pixel constraints of \citet{mcgreer15}.  One of the main motivations of this work is to explore whether LAE measurements could provide a complementary test of such a model.

 The other parameters were fixed by matching to observational constraints provided by SILVERRUSH and quasar absorption spectra measurements.   To fix the LAE parameters, we considered the mean LAE surface density, REW distribution, Ly$\alpha$ luminosity function, and 2PCF.  Most importantly for the ensuing discussions, at a given redshift, all of our models were calibrated to have approximately the same mean LAE surface density, as shown in the last column of Table \ref{tab:beta_coeff}.  To achieve this, we adjusted the proportionality between halo circular velocity and line offset ($\beta$) to match our models approximately to SILVERRUSH surface densities. Our model surface densities are closest to those of the ``Homogeneous sample" described in \citet{2018PASJ...70S..13O}.\footnote{For reference, the Homogeneous sample from \citet{2018PASJ...70S..13O} has surface densities of 70 and 41 deg$^{-2}$ at $z=5.7$ and 6.6, respectively. }   We found that a reasonable match could be obtained for all the other LAE constraints by fixing the duty cycle to $\Delta t = 50$ Myr.  

The mean ionizing intensity in our fluctuating UVB models (\S \ref{sec:ESMR}) were calibrated by Ly$\alpha$ forest measurements where available.  At $z=5.7$, we constructed synthetic Ly$\alpha$ forest sight lines through our simulations and rescaled the intensity, assuming photoionization equilibrium, to match the mean flux measurement of \citet{2018MNRAS.479.1055B}.  Currently, forest segments at $z=6.6$ are too saturated and sparse to provide a reliable estimate of the mean flux.  For simplicity, we held the mean intensities fixed with redshift between the $x_{\rm HI} = 0$ models, the $x_{\rm HI} = 0.1$ and 0.25 models, and the $x_{\rm HI} = 0.3$ and 0.5 models. This can be seen in the mean hydrogen photoionization rates in ionized regions provided in Table \ref{tab:beta_coeff}.  We tested models in which $\langle \Gamma_{\rm HI} \rangle$ is a factor of 5 lower at $z=6.6$ than at $z=5.7$.  We found that $\langle\Gamma_{\rm HI}\rangle$ is almost completely degenerate with $\beta$, motivating our choice to hold the former fixed with redshift.   The mean free path through ionized gas, $\lambda^{\rm mfp}_{912}$, is another key parameter in our UVB model. We fixed this parameter by extrapolating to higher redshift the empirical fit of \citet{2014MNRAS.445.1745W}.  We note that $\lambda^{\rm mfp}$ during reionization could be considerably shorter than the values adopted here owing to the potentially lower $\Gamma_{\rm HI}$ as well as relaxation effects \citep{2020ApJ...898..149D}.

Figure \ref{fig:obs_anchors_57} shows that our models are in reasonable agreement with measurements of the REW distribution (left), luminosity function (middle), and 2PCF (right) from SILVERRUSH observations.  The top and bottom rows correspond to $z=5.7$ and $6.6$, respectively. In all top (bottom) panels, the red, blue, and green curves respectively correspond to $\bar{x}_{\rm HI}=0.0 ~(0.0)$, $0.1~(0.25)$, and $0.3~(0.5)$.  Additionally, the grey curves show the intrinsic statistics before attenuation by the IGM.  For comparison, the data points show some of the most recent observational measurements. IGM opacity acts to skew the REW distribution toward lower values.  It also changes the normalization and shape of the luminosity function, and increases the observed clustering of LAEs.  These trends are consistent with previous findings \citep[e.g.][]{2007MNRAS.381...75M, 2014MNRAS.444.2114J, 2019MNRAS.485.1350W}.  Similar to the results of \citet{2019MNRAS.485.1350W}, our 2PCFs exhibit more clustering compared to the SILVERRUSH measurements at $z=6.6$. This may owe to the relatively massive halos that our LAEs inhabit.  Our models have $M_{\rm min} = 3 \times 10^{10}~ h^{-1}$ M$_{\odot}$ and an average mass of $\langle {\rm M} \rangle= 1.5\times 10^{11}~h^{-1}{\rm M_{\odot}}$.  In contrast, SILVERRUSH reports $M_{\rm min} \sim 2.1\times 10^{9}  (8.8 \times 10^{8} )~h^{-1}{\rm M_{\odot}}$, and average masses of $\sim 8.7\times 10^{10}(4.4\times 10^{10})~h^{-1}{\rm M_{\odot}}$ at z=5.7(6.6) \citep{2018PASJ...70S..13O}.  These masses were derived by applying the halo occupation distribution (HOD) modelling of \citet{2016ApJ...821..123H} to the measured 2PCF.  Another effect at play could be the lower surface densities in our models compared to the Homogenous sample used to measure the 2PCF.

\begin{figure*}
    \centering
    \includegraphics[width=15cm]{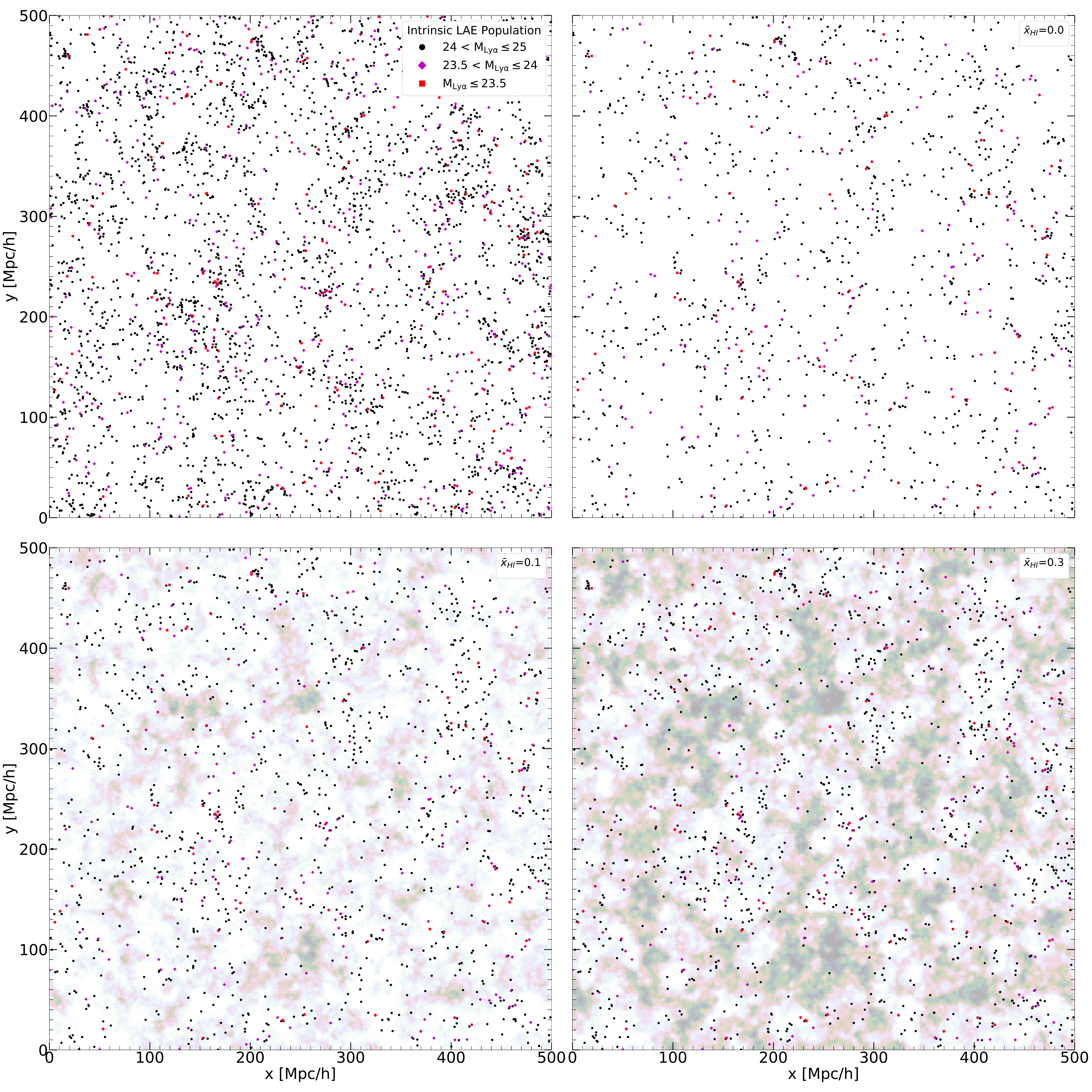}
    \caption{Example $500\times500$ ($h^{-1}$Mpc)$^2$ slices through our simulated LAE distributions at $z=5.7$.  The different dot-colors correspond to the Ly$\alpha$ apparent narrow-band magnitude ($M_{\mathrm{Ly} \alpha}$) ranges shown in the top-left panel.  The top-left panel shows the intrinsic LAE sample before any attenuation by the IGM is applied.  The remaining panels correspond to our models with $\bar{x}_\mathrm{HI}=0.0$ (top-right), $\bar{x}_\mathrm{HI}=0.1$ (bottom-left), and $\bar{x}_\mathrm{HI}=0.3$ (bottom-right).  The shading shows the integrated neutral fraction along the line of sight, which is $30~h^{-1}$Mpc deep.  Darker shades indicate regions with higher neutral fractions.  These three panels have approximately the same mean LAE surface density, $\langle \Sigma \rangle \approx 72$ (deg)$^{-2}$.  At fixed $\langle \Sigma \rangle$, a higher neutral fraction obscures LAEs in the voids, but these must be compensated by more visible LAEs elsewhere.  Hence the presence of neutral gas leads to more apparent clustering of the LAEs. Notably, the neutral gas is relegated to the voids in the tail end of reionization, suggesting that LAE void statistics may provide an informative test of models in this regime.          }
    \label{fig:ex_slice_57}
\end{figure*}

\begin{figure*}
    \centering
    \includegraphics[width=15cm]{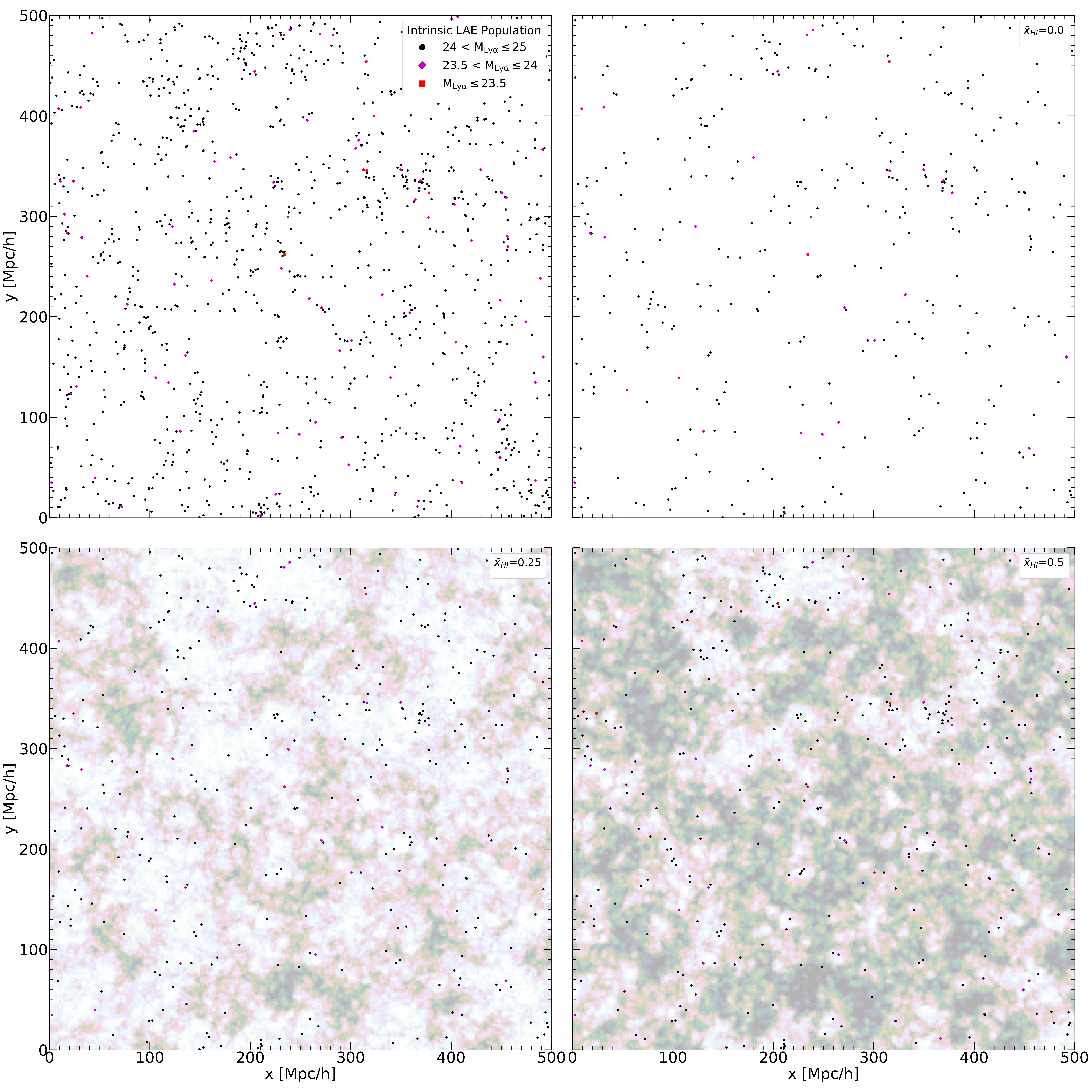}
    \caption{Same as Figure \ref{fig:ex_slice_57}, except at $z=6.6$.  The top-right ($\bar{x}_\mathrm{HI}=0.0$), bottom-left ($\bar{x}_\mathrm{HI}=0.25$), and bottom-right  ($\bar{x}_\mathrm{HI}=0.5$) panels have a fixed $\langle \Sigma \rangle \approx 32$ (deg)$^{-2}$. }
    \label{fig:ex_slice_66}
\end{figure*}

Figures \ref{fig:ex_slice_57} and \ref{fig:ex_slice_66} show mock survey fields of size $(500h^{-1}\mathrm{Mpc})^2$ at $z=5.7$ and $6.6$, respectively.  For all panels we impose the survey cuts described in \S \ref{sec:mock_surveys}.  The top-left panels show the intrinsic distribution of LAEs in the slices, i.e. without attenuation by the IGM.   The subsequent plots (moving clockwise) correspond to increasing global neutral fractions, where the shading shows the projected neutral gas density.    The different dot-colors correspond to Ly$\alpha$ magnitude bins, as denoted in the plot legends.  In the latter half of reionization, the neutral gas is increasingly confined to voids in the LAE distribution.  This gas attenuates the Ly$\alpha$ lines of background LAEs, obscuring some of them from view. Importantly, because we fix the surface density of LAEs to be approximately the same between our models, these missing LAEs are compensated by an increased density of LAEs in the ionized regions. This effect contributes to the enhanced clustering seen in the 2PCFs (Fig. \ref{fig:obs_anchors_57}), and we will see its consequences for peak statistics later in this paper.    

\section{Results}
\label{sec:stats}

\subsection{Cosmic Variance}
\label{sec:cos_var}

We begin with a broad comparison of our simulated ensembles to the SILVERRUSH fields.  
 We have applied homogeneous luminosity and color cuts to the publicly available SILVERRUSH catalogs (see text in \S \ref{sec:surf_dens}).  The LAE surface density of each field is reported in Table \ref{tab:sig_dsig_table}.   The data exhibits significant field-to-field variation. At $z=5.7$, D-ELAISN1 has a surface density of just $\approx 31$ deg$^{-2}$, whereas UD-SXDS is a factor of $3$ denser with $\approx 98$ deg$^{-2}$.   Additionally, some fields, such as D-ELAISN1 at z=5.7 and D-DEEP2-3 at z=6.6, show conspicuous asymmetries over $\sim 200~h^{-1}$Mpc scales (see e.g. Figs. 4 and 5 of \citealt{2018PASJ...70S..13O}). Some regions remain largely empty of LAEs, while others contain significant over-densities.  In this section, we explore whether the observed variations are consistent with expectations from large-scale structure formation alone, or if some other source of variance is required. We also quantify the contribution of reionization to the expected cosmic variance.

\subsubsection{LAE Surface Density}
\label{sec:surf_dens}

\begin{figure}
    \includegraphics[width=8.25cm]{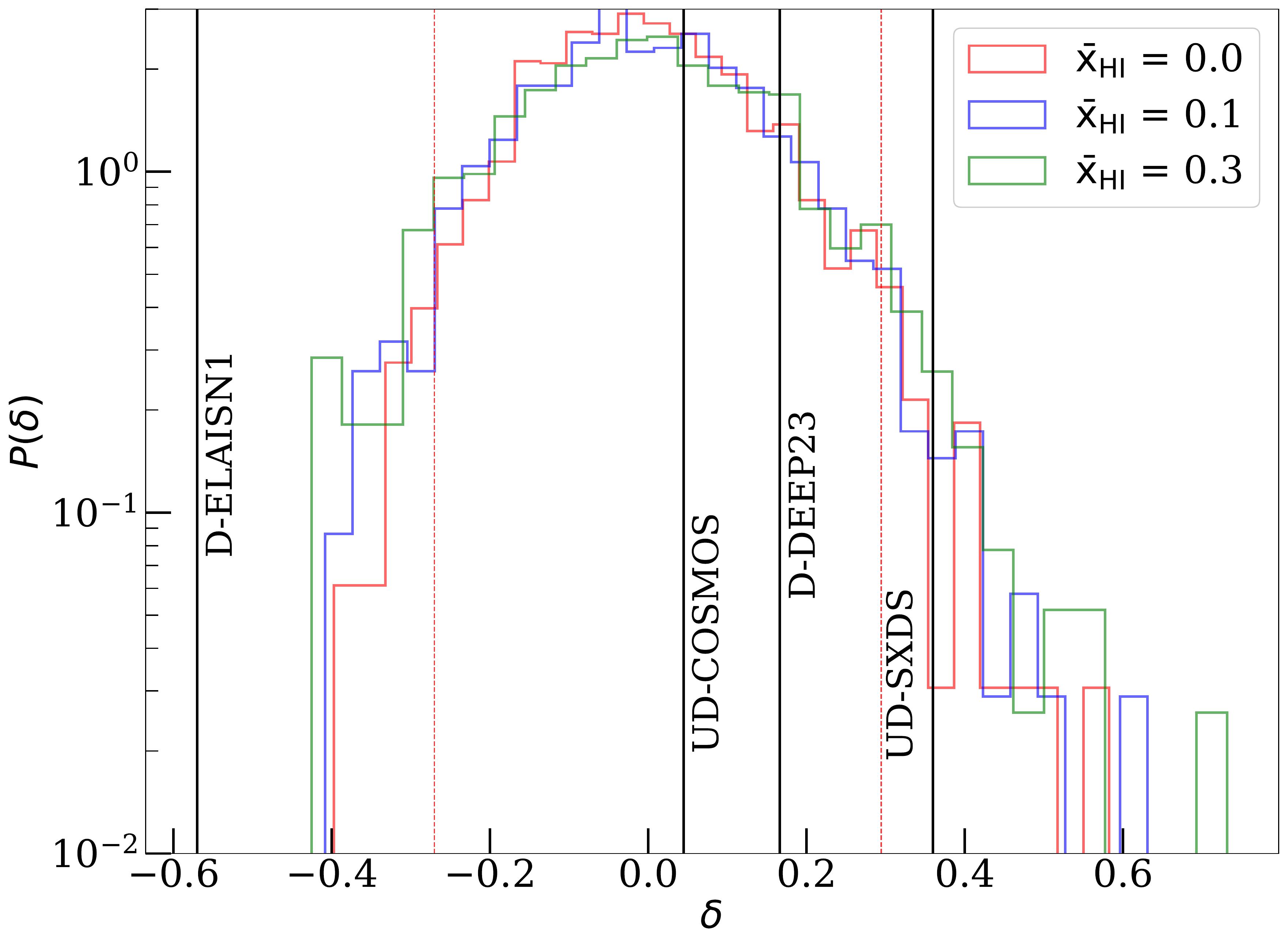}\hfill
    \includegraphics[width=8.25cm]{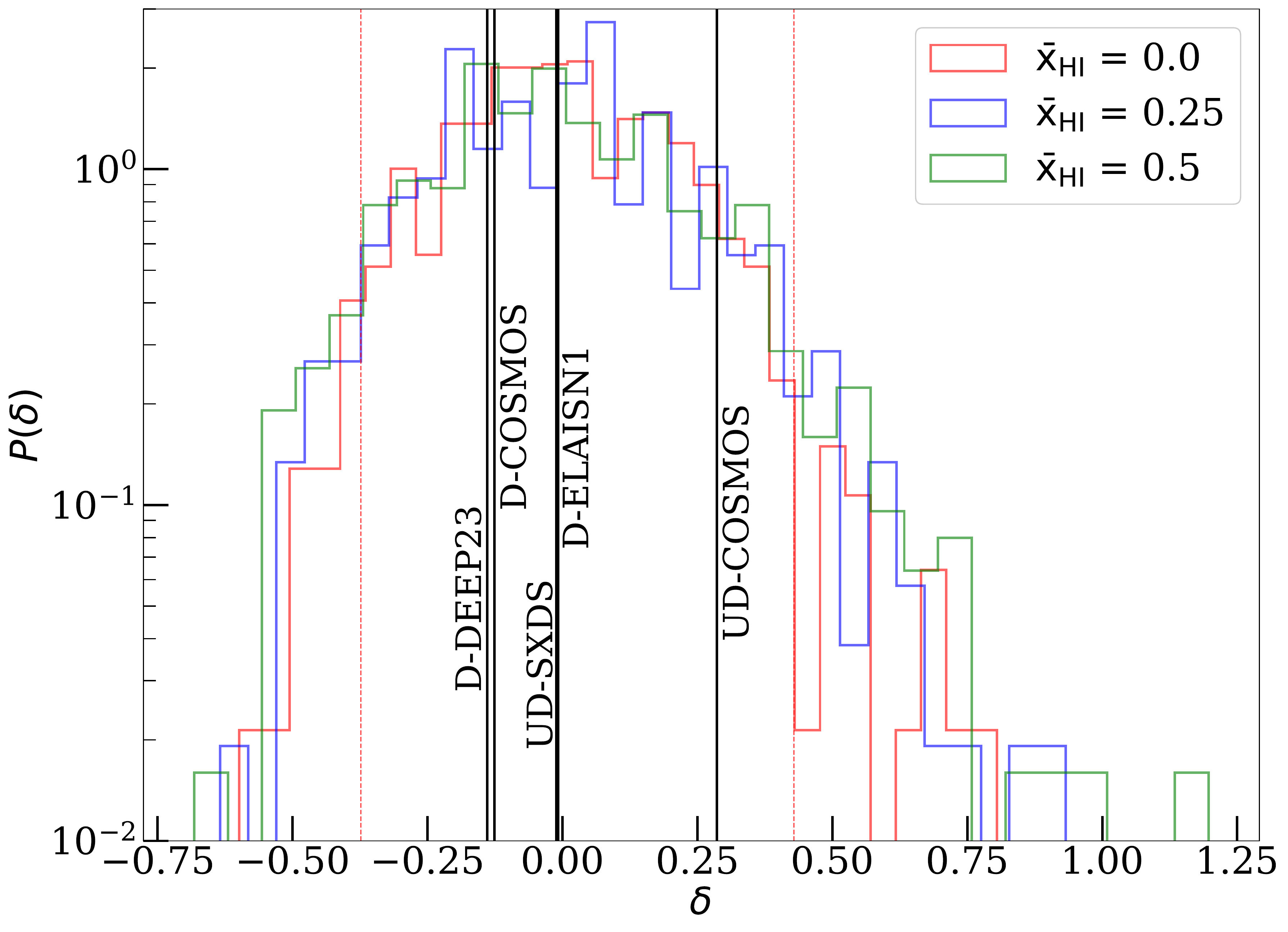}\hfill
    \caption{Cosmic variance in the surface density of SILVERRUSH-sized fields of view.  Here we show the probability distribution of the surface density contrast, $\delta=\Sigma/\langle \Sigma \rangle - 1$.  The top and bottom panels correspond to $z=5.7$ and $z=6.6$, respectively.  The red, blue, and green histograms represent our models with different global neutral fractions.  The vertical black lines represent $\delta$ measured from the publicly available SILVERRUSH catalogs \citep{2018PASJ...70S..14S}, with uniform luminosity and color cuts applied.   The red, thin-dashed lines show the $2\sigma$ range of our surface density distribution in a completely ionized scenario.  Black lines are annotated with individual field names. A more neutral IGM yields only a slight increase in the cosmic variance. Our model distributions are consistent with the data, aside from D-ELAISN1, which is likely affected by poor seeing \citep{2018PASJ...70S..16K}.}
    \label{fig:sigm_distr}
\end{figure}

In Figure \ref{fig:sigm_distr} we show the LAE surface density distributions in our reionization models.  The top and bottom panels correspond to $z=5.7$ and $z=6.6$, respectively. The red, blue and green histograms represent different neutral fractions, as denoted in the plot legends. For each model, the distributions consist of $1,000$ mock FoVs. To facilitate comparisons with the observations, we have matched the sizes of our fields to the mean sizes of the SILVERRUSH fields (see \S \ref{sec:mock_surveys}).  This is not ideal because the SILVERRUSH fields vary in size, from 2-6 square degrees \citep{2018PASJ...70S..14S, 2018PASJ...70S..16K}.  However, we will find that the observed cosmic variance can be reproduced reasonably with our simplistic approach.  To remove the effect of small differences in the mean surface densities in our models, we plot the surface density contrast, $\delta = (\Sigma- \langle \Sigma \rangle)/\langle \Sigma \rangle$, where $\langle \Sigma \rangle$ is the average surface density.  Figure \ref{fig:sigm_distr} shows that reionization increases the cosmic variance in $\delta$ quite subtly.  For reference, the standard deviations of $\delta$ between our models are $0.14(0.20)$, $0.15(0.23$), and $0.17(0.25)$ for neutral fractions of $\bar{x}_{\rm HI} = 0.0(0.0)$, $0.1(0.25)$, and $0.3(0.5)$ at $z=5.7(6.6)$, respectively.
 
The black vertical lines in Figure \ref{fig:sigm_distr} illustrate the cosmic variance in the SILVERRUSH data. Each line is annotated with its respective field name (see Table \ref{tab:sig_dsig_table}). The luminosity and color cuts applied in the published SILVERRUSH analyses vary, depending on the application.  Hence, we downloaded the public SILVERRUSH catalogs\footnote{\url{http://cos.icrr.u-tokyo.ac.jp/rush.html}} and applied our own cuts with the goal of matching our simulated catalogs to the extent possible. Specifically, at $z=5.7$, we selected Ly$\alpha$ luminosities $> 6.3\times 10^{42}$ ${\rm erg/s}$ and applied a color cut of $i-$NB816 $>1.2$. This yields $\langle \Sigma \rangle=72$ deg$^{-2}$, in agreement with our models (see Table \ref{tab:beta_coeff}).  At $z=6.6$, we used $>1.41\times 10^{43}$ ${\rm erg/s}$ and $z-$NB921$>1.8$.  The higher luminosity limit adopted for $z=6.6$ ensures that a homogeneous cut can be applied across all fields in the public SILVERRUSH data.  Note that this limit is much higher than in our fiducial $z=6.6$ catalog used throughout the rest of this paper, and results in our models having $\langle \Sigma \rangle = 9.5$ deg$^{-2}$ at $z=6.6$.  The color cuts, which are from \citet{2018PASJ...70S..16K}, roughly match the REW cuts applied to our simulated catalogs.  For comparison, the red/dashed vertical lines show the 2$\sigma$ limits of the model distributions with $\bar{ x}_{\rm HI}= 0$. (The 2$\sigma$ limits for the other models are similar.) 

We find that the observed field-to-field variation is almost entirely consistent with expectations from large scale structure alone, and that reionization is unlikely to play a significant role in increasing this variation.  At $z=5.7$, our models, however, cannot account for the dearth of LAEs observed in D-ELAISN1.   However, as noted by \citet{2018PASJ...70S..16K}, the deficit may owe to poor seeing for that field. Interestingly, D-ELAISN1 also shows a strong asymmetry in the LAE distribution, although there is no evidence of spatial variations in the data quality \citep{2018PASJ...70S..13O}.  In the next section we explore the nature of such asymmetries.

\subsubsection{Split-screen contrast}
\label{sec:ssv}

We quantify half-plane variations in the LAE fields using a statistic that we term the ``split-screen contrast."  The FoV is split, either vertically, horizontally, or diagonally to create regions of roughly equal area.  We then calculate the contrast, 
\begin{equation}
    \frac{\Delta \Sigma}{ \Sigma} =  \frac{ |  \Sigma_1 - \Sigma_2 |}{ \Sigma }
    \label{SSV}
\end{equation}
where $\Sigma_1$ and $\Sigma_2$ are the surface densities of the two halves (or ``split screens") and $ \Sigma$ is the surface density of the field.

\begin{figure}
    \subfloat{%
    \includegraphics[width=8.25cm]{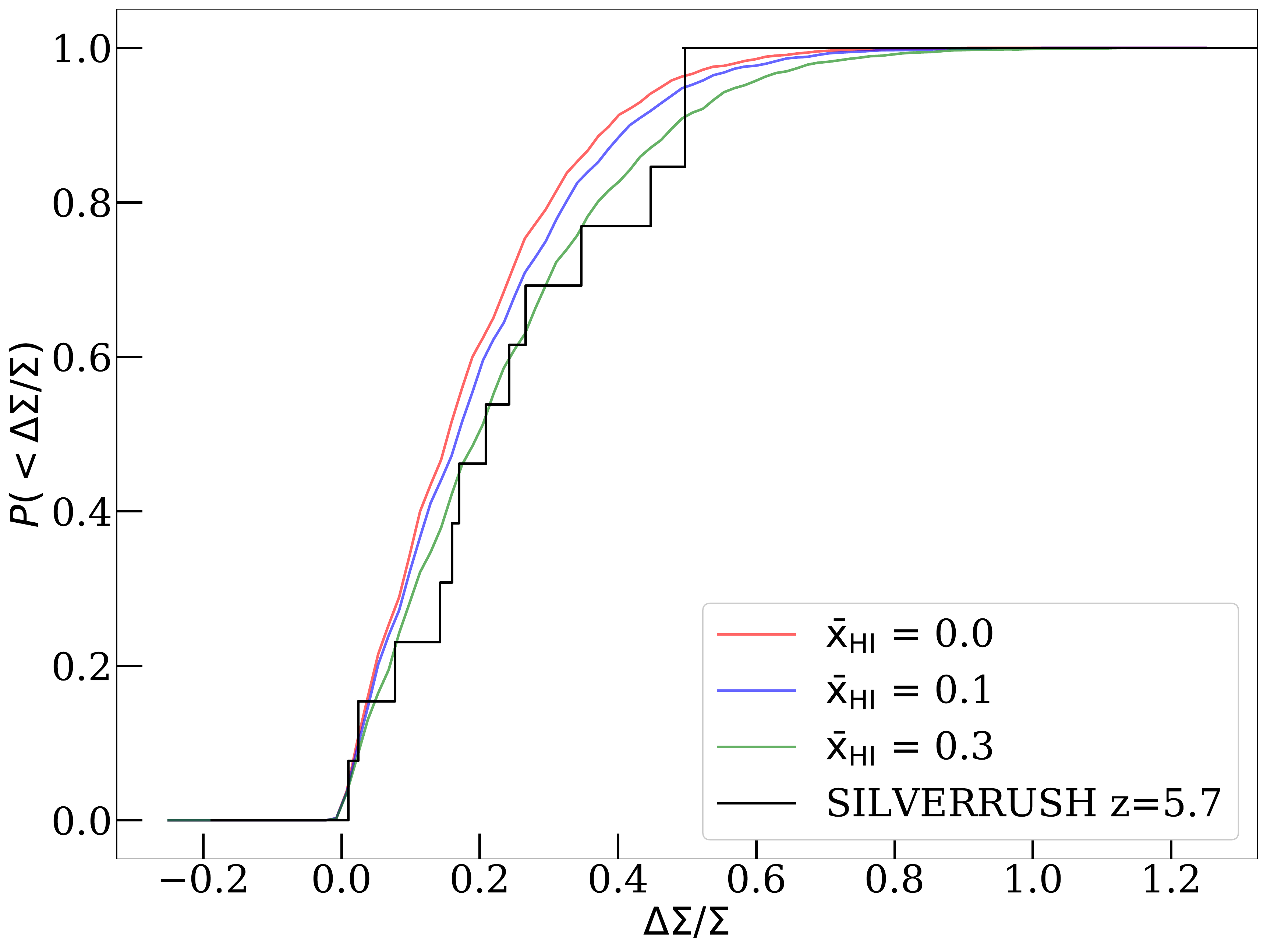}}\hfill
    \subfloat{%
    \includegraphics[width=8.25cm]{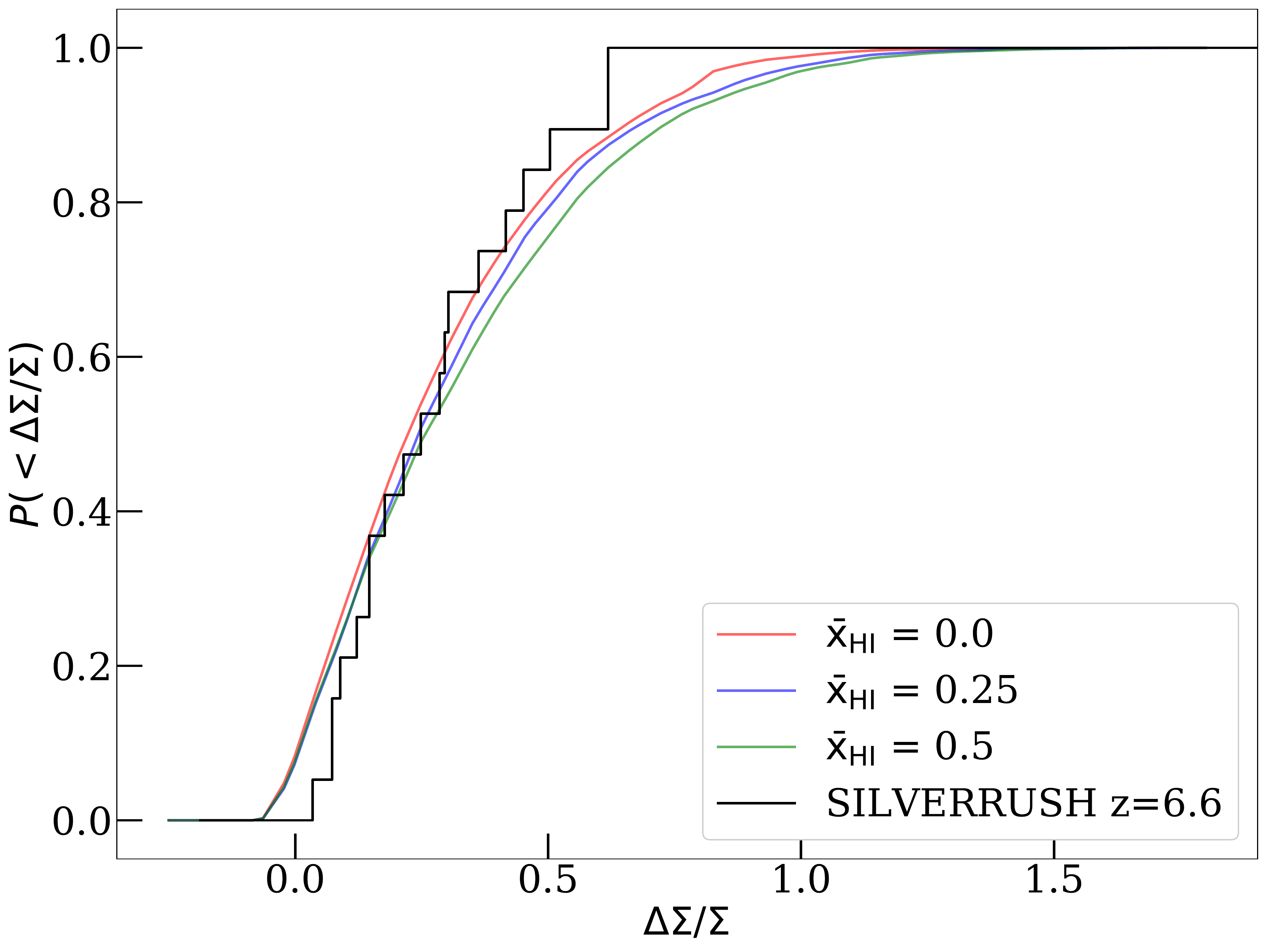}}\hfill
    \caption{Cumulative probability distribution of the split-screen contrast, $\Delta \Sigma / \Sigma$.  The split-screen contrast quantifies half-plane asymmetry in the LAE distribution (see main text for more details).  The top and bottom panels correspond to $z=5.7$ and $z=6.6$, respectively. The colored curves correspond to our reionization models. The black histograms show measurements from the publicly available SILVERRUSH catalogs \citep{2018PASJ...70S..14S}, with uniform luminosity and color cuts applied.  The likelihood of finding a field with high $\Delta \Sigma / \Sigma$ increases slightly with neutral gas fraction. However, the visually apparent half-plane asymmetries seen in some of the SILVERRUSH fields are consistent with expectations from large-scale structure formation alone. }
    \label{fig:dsig_distr}
\end{figure}

In Figure \ref{fig:dsig_distr} we show the cumulative probability distribution of split-screen contrasts in our models, $P(<\Delta \Sigma/ \Sigma$), or the probability of finding $\Delta \Sigma/\Sigma$ less than a given value.  Reionization widens the distribution somewhat.  At $z=5.7$, the standard deviation increases by 30\% between a completely ionized IGM and $\bar{x}_{\rm HI} = 0.3$.  At $z=6.6$ the differences are more subtle, a 12\% increase in width between $\bar{x}_{\rm HI} = 0.0$ and $0.5$, owing to the relative dominance of shot noise from a lower mean surface density.  Generally speaking, a larger neutral fraction extends the high-contrast tail of the distribution, enhancing the likelihood of finding more extreme half-plane asymmetries.

For comparison, we attempt to quantify the split-screen contrasts in the SILVERRUSH data using the same luminosity and color cuts described in the last section.  We note that the fields have different sizes and non-trivial shapes.  To account for this, we inspected the fields and found axes of symmetry along which to perform the bisections, maintaining roughly equal areas between the split screens.  The black histograms in Figure \ref{fig:dsig_distr} show the results of our crude measurements, which are reported in Table \ref{tab:sig_dsig_table}. (Individual fields have more than one split-screen configuration.)   The largest contrasts are found in D-ELAISN1 at $z=5.7$, and in D-COSMOS and UD-COSMOS at $z=6.6$. Note that we have applied our own homogeneous luminosity and color cuts to the data, so the split screen contrasts that we report here may not be visually apparent in published visualizations of the fields \citep[e.g.][]{2018PASJ...70S..13O}.   

Our measurements from SILVERRUSH lie comfortably within the dispersion of our models.  A Kolmogorov–Smirnov test indicates that the observed distributions are consistent with all of the models. For example, at $z=5.7$, we find p-values of 0.8 and 0.3 for $\bar{x}_{\rm HI} = 0.3$ and 0.0, respectively. We conclude that the half-plane asymmetries present in the SILVERRUSH data are consistent with the expected large scale structure alone.  Patchy reionization does, however, enhance the probability of observing larger split-screen contrasts.  For example, the probabilities of finding $\Delta \Sigma/ \langle \Sigma \rangle \ge 0.5$ are 4(21)\%, 6(23)\%, 10(27)\% in our models with $\bar{x}_{\rm HI} = 0.0 (0.0)$, $0.1(0.3)$, and $0.3(0.5)$ at z=5.7(6.6), respectively.

In conclusion, we have found that our ensembles of simulated fields are consistent with the cosmic variance and half-plane asymmetries observed in the SILVERRUSH data.   

\begin{table}
    \begin{tabular}{|c|c|c|c|}
  z & Field & $\Sigma$ [deg$^{-2}$] & $\Delta \Sigma$ /$ \Sigma$ \\ \hline 
    5.7 & UD-COSMOS & 75.6 & 0.27, 0.14\\ \hline 
    & UD-SXDS       & 98.4 & 0.33, 0.03 \\ \hline
    & D-DEEP2-3     & 84.4 & 0.28, 0.46, 0.18 \\ \hline
    & D-ELAISN1     & 31.1 &  0.20, 0.18, 0.76, 0.02, 0.55, 0.58 \\\hline \hline
    6.6 & UD-COSMOS & 12.2 & 0.75, 0.32 \\ \hline 
    & UD-SXDS       & 9.41 & 0.11, 0.55 \\ \hline
    & D-COSMOS      & 8.27 & 0.75, 0.44, 0.31\\ \hline
    & D-DEEP2-3     & 8.16 & 0.31, 0.15, 0.08, 0.15, 0.23, 0.08 \\ \hline
    & D-ELAISN1     & 9.38& 0.15, 0.07, 0.44, 0.22, 0.29,  0.51  \\ \hline
    \end{tabular}
    \caption{Surface densities (3rd column) and split-screen contrasts (4th column) measured from our sample of the SILVERRUSH data.  Our luminosity and color cuts are described in \S \ref{sec:surf_dens}. This data is compared against our models in Figures \ref{fig:sigm_distr} and \ref{fig:dsig_distr}.  } 
    \label{tab:sig_dsig_table}
\end{table}

\subsection{Voids}
\label{sec:voids}

\begin{figure}
    \vspace{0.cm}
    \includegraphics[width=8.6cm]{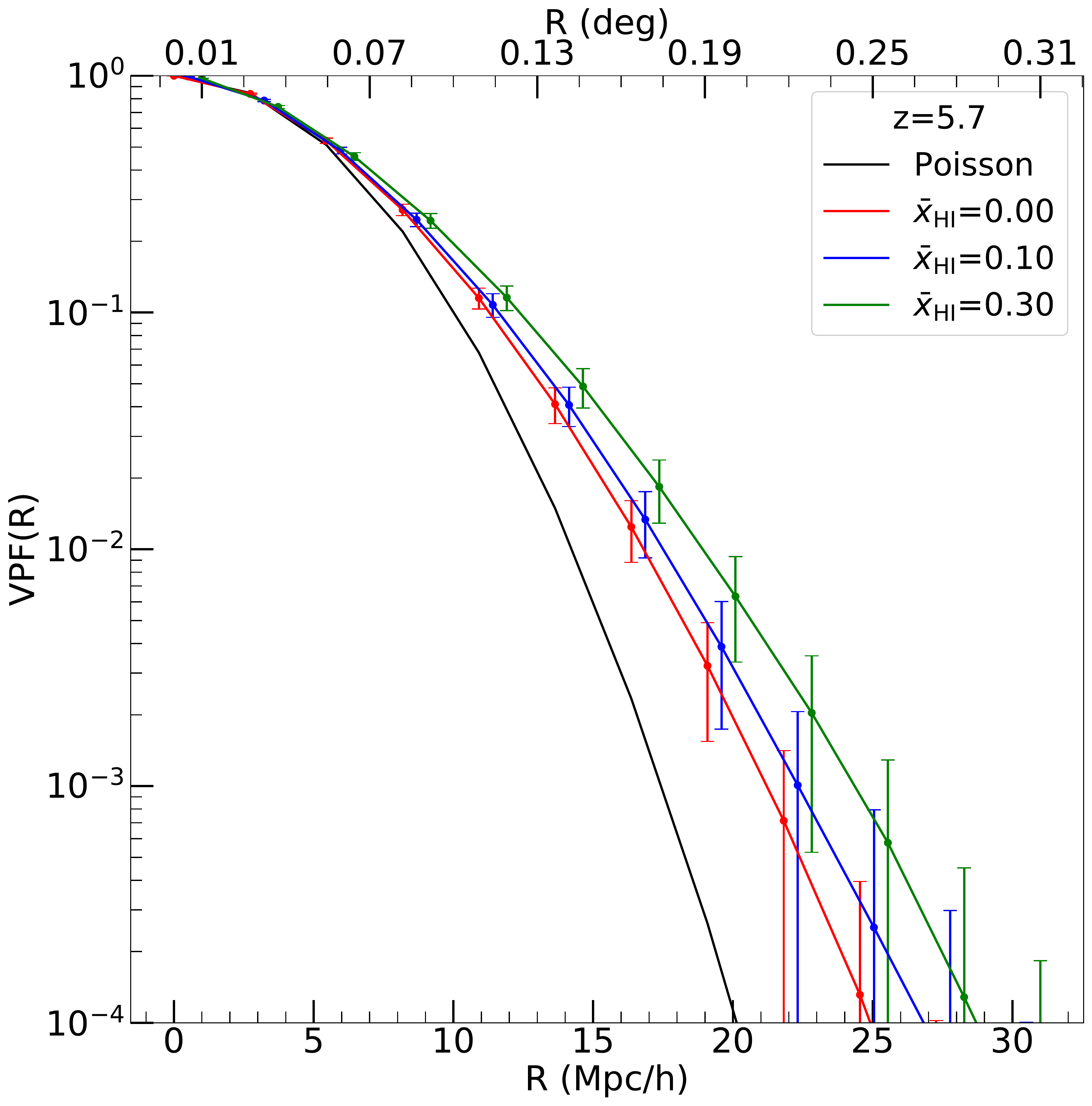}
    \vspace{0.cm}
    \includegraphics[width=8.6cm]{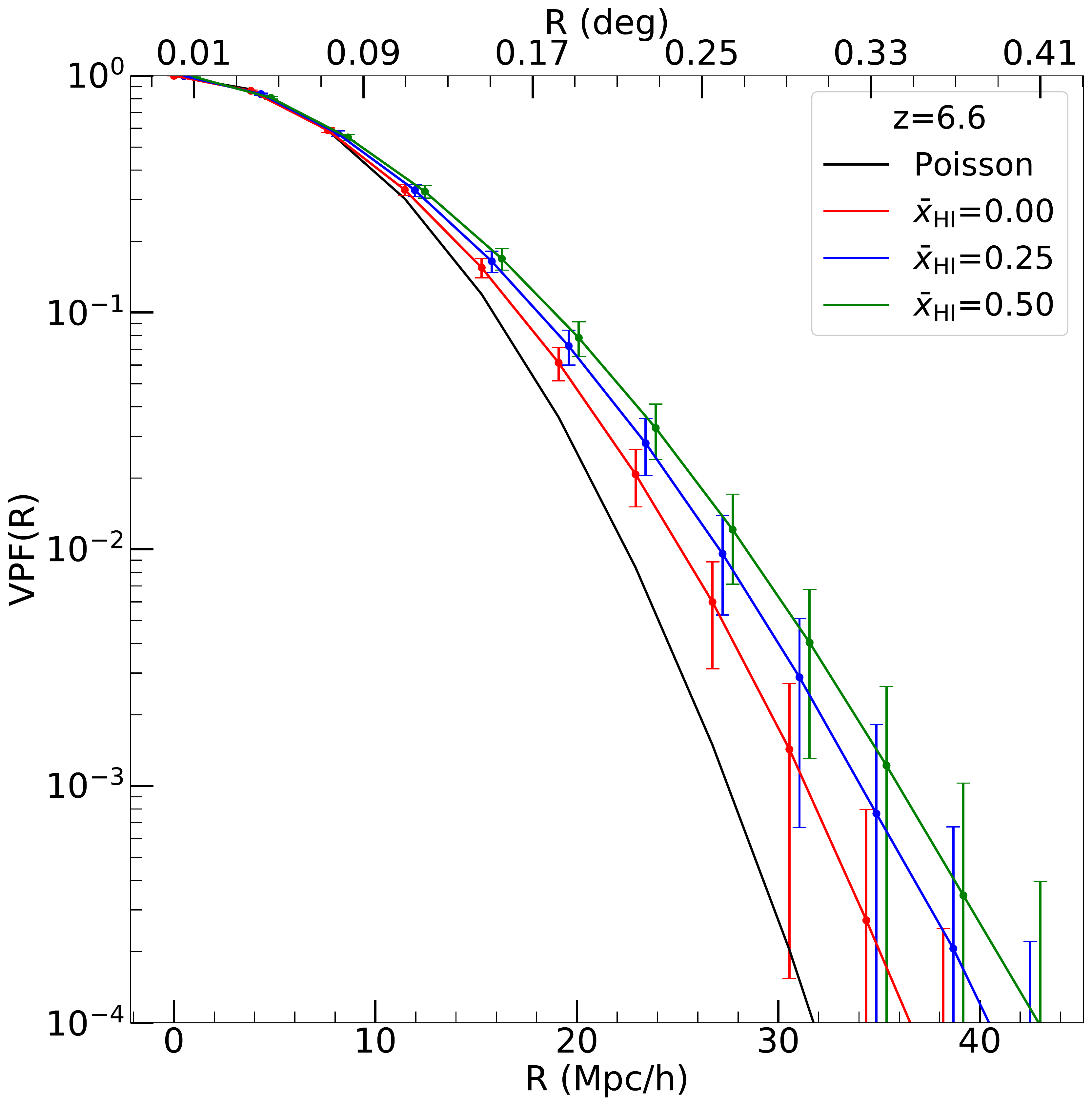}\hfill
    \caption{The void probability function (VPF) at $z=5.7$ (top) and $z=6.6$ (bottom) for our models with different global neutral fractions.  The black curves show the analytic result for a uniform-random distribution of LAEs with matched surface density.  The error bars show the $1\sigma$ dispersion from cosmic variance in a 10-field survey, or a SILVERRUSH-like survey with roughly double the size.  A larger neutral fraction increases the probability of finding larger empty circles.  This is a consequence of LAEs being obscured in the low-density regions of the IGM, which are the last places to be reionized.}
    \label{fig:vpf_boot}
\end{figure}

Clustering statistics of LAEs provide an important observational window into the reionization process.   Previous studies have focused on using the 2PCF to constrain the neutral fraction of the IGM \citep{2007MNRAS.381...75M, 2010ApJ...723..869O, 2014MNRAS.444.2114J, 2015MNRAS.450.4025H, 2015arXiv150502787S, 2018PASJ...70S..13O}. However, Figure \ref{fig:ex_slice_57} illustrates that, toward the end of reionization, neutral gas is mostly relegated to voids in the LAE distribution.  This suggests that void statistics could be more informative in the final stages of reionization, when the neutral gas disproportionately obscures LAEs in under-dense regions. Also working in favor of this idea is the fact that the space density of $z=5.7$ LAEs is considerably larger than at e.g. $z=6.6$.   In what follows, we examine the prospect of constraining reionization with the statistics of voids.

We quantify reionization's effect on LAE void statistics using the void probability function (VPF), or the probability of finding a circle of radius R containing no LAEs (\citealt{2006ApJ...647..737T}; \citealt{2008ApJ...686...53T}; \citealt{2007MNRAS.381...75M}; Perez et al. in prep.). The VPF is complementary to the 2PCF, encoding correlations at orders beyond the two-point function.  A key advantage of the VPF is that it is simple and can be applied even to sparse LAE fields, although see Perez et al. in prep. for important caveats on its application.  We compute the VPF by randomly placing down circles with radius R in our mock fields and then finding the fraction of circles that do not contain LAEs.

In Figure \ref{fig:vpf_boot} we show our VPF results at $z=5.7$ (top) and $z=6.6$ (bottom). The red, blue, and green curves show our reionization scenarios as denoted in the plot legends. For comparison, the black curves show the VPF in the case of a uniform-random distribution of LAEs, $P(R) = \exp(-\pi R^2 \langle \Sigma \rangle )$, with the same mean surface density (see Table \ref{tab:beta_coeff} for $\langle \Sigma \rangle$).  A comparison between the black and red curves shows that large scale structure clearly enhances the VPF above the Poisson expectation.  Consistent with previous works, we find that an ongoing reionization process further enhances the VPF. Neutral gas in the cosmic voids attenuates Ly$\alpha$ emission, making holes in the LAE distribution appear larger (cf. Figs. \ref{fig:ex_slice_57} and \ref{fig:ex_slice_66}).  A higher global neutral fraction enhances the probability of finding empty circles on larger scales. At $z$=6.6 we find that the probability of finding larger empty circles is enhanced significantly. (Note the different scales in the panels.)  This owes to two effects.  First, there are simply fewer LAEs at $z=6.6$, which results in larger random gaps in the distribution (see \citealt{2007MNRAS.382..860D} for a discussion). Second, the larger neutral fraction results in more LAEs being obscured.  

Even on the $L\sim 200~h^{-1}$ Mpc scales of our fields, large-scale power can lead to substantial cosmic variance in our void statistics. Such variations can pose a challenge for constraining reionization with voids.   We quantify the cosmic variance by bootstrap resampling our simulated FoV ensemble. If our mock survey contains $N$ fields, we draw $N$ samples with replacement and calculate the VPF from these samples. The resulting VPF constitutes one realization of a mock survey, and we generate $10,000$ of these. We have checked that our results are well converged in the number of realizations.  In what follows, we consider two survey configurations; (1) The current SILVERRUSH size: $N=4$ and $5$ for $z=5.7$ and $6.6$, respectively; (2)  A future SILVERRUSH-like survey covering twice the area, $N=10$.  The error bars in Figure \ref{fig:vpf_boot} show the $1\sigma$ dispersion from cosmic variance in the VPF assuming the futuristic $N=10$ field survey.  From here on, all cosmic variance error bars assume $N=10$.

Can the VPF provide stronger constraints than the 2PCF on the tail end of reionization, as we have argued above?  We address this question with a simple $\chi^2$ analysis. We gauge the significance at which a statistic can rule out our model with $\bar{x}_{\rm HI}= 0.3~(0.5)$ at $z=5.7~(6.6)$, assuming that the true model is $\bar{x}_{\rm HI}= 0$.  To this end, we used our bootstrap samples of the $\bar{x}_{\rm HI}= 0$ model to compute the covariance matrix\footnote{We have checked that our calculation is well converged in number of realizations.}, taking into account correlations between radial bins.  We then calculated $\chi^2 = \boldsymbol{P}^{T} C^{-1}\boldsymbol{P}$, where $C$ is the covariance matrix and $\boldsymbol{P}$ is the difference vector between the $0.3 (0.5)$ model and a mock survey drawn from the $\bar{x}_{\rm HI}= 0$ model. Finally, we calculated p-values to assess the rarity of the obtained $\chi^2$. Given the non-Gaussianity exhibited in some of the radial bins, we integrated the numerical $\chi^2$-distribution from the bootstrapping to compute the p-values.

A summary of our simple $\chi^2$ analysis is reported in Table \ref{tab:pval_table}.  There we show p-values for the 2PCF, VPF, as well as two of the peak-based statistics that we explore in the next section. To ensure that our conclusions are not biased by a random fluctuation, the p-values reported there are the average values over $1,000$ mock observations.  Results are given for current and ``futuristic" (10-field) survey configurations. We note that our analysis adopts cosmic variance error bars, ignoring all other observational uncertainties. The numbers in parentheses correspond to cases in which we assume that a fraction of the LAEs are actually misidentified foreground interlopers. (See discussion below.) Let us consider first the ideal case with no contamination. We find that the VPF outperforms the 2PCF at testing late reionization models in which the tail end extends below $z=5.7$. On the other hand, for both survey configurations, the 2PCF and VPF provide similar constraining power for a model with $\bar{x}_{\rm HI} = 0.5$ at $z=6.6$.   These results have an intuitive interpretation. The neutral gas is mostly in the voids at $z=5.7$, leaving a stronger signature in the VPF compared to the 2PCF.

\begin{table}
    \begin{tabular}{|c|c|c|c|c|c|c|}
   &  & p-value & p-value \\
    z& Statistic& (SILVERRUSH-like)&  (10 fields)& \\\hline \hline 
    5.7 &  2PCF	& 0.26 	& 0.18\\ 
    &  10\% contamination		    & (0.28)	& (0.22) \\
    &  	30\% contamination	    & (0.34)	& (0.32) \\ \hline
    &  VPF		& 0.18	    & 0.07 \\ 
    &  10\% contamination		    & (0.23)	& (0.12) \\
    &  	30\% contamination	    & (0.31)	& (0.23) \\\hline
    &  PTF 		& 0.42 	    & 0.30   \\ \hline
    &  PPF 		& 0.63 	    & 0.51 	\\ \hline \hline
    6.6 &  2PCF	& 0.19 	& 0.08\\ 
    &  	10\% contamination	    & (0.26)    & (0.12) \\
    &  	30\% contamination	    & (0.32)	& (0.21) \\\hline
    &  VPF		& 0.21 	    & 0.09 	\\ 
    &   10\% contamination        & (0.25)    & (0.14)   \\ 
    &    30\% contamination       & (0.34)    & (0.23)   \\\hline
    &  PTF 		& 0.19 	    & 0.10 	\\ \hline
    &  PPF 		& 0.58 	    & 0.25 	\\ \hline
    \end{tabular}
    \caption{Simple $\chi^2$ analysis to gauge the efficacy of void and peak statistics for testing reionization models. The reference models have $\bar{x}_{\rm HI} = 0$ and we compute the probabilities of obtaining $\chi^2$ at least as extreme as those of the $\bar{x}_{\rm HI} = 0.3$ and 0.5 models at $z=5.7$ and $6.6$, respectively, assuming cosmic variance uncertainties. These p-values are shown for two survey configurations: (1) A SILVERRUSH-like survey with 4 and 5 fields at $z=5.7$ and $z=6.6$, respectively; (2) An expanded SILVERRUSH-like survey with 10 fields (roughly doubled in area). We consider the two-point correlation function (2PCF), Void Probability Function (VPF; \S \ref{sec:voids}), Pixel Threshold Fraction (PTF; \S \ref{sec:ptf}), and Peak Probability Function (PPF; \S \ref{sec:ppf}).  The numbers shown in parentheses for the 2PCF and VPF assume that 10\% or 30\% of the LAEs are misidentified foreground sources. All other numbers assume no contamination.  The VPF is particularly well-suited for probing the tail end of reionization, where LAE surface densities are highest.   }
    \label{tab:pval_table}
\end{table}

In the absence of full spectroscopic confirmation, however, narrow band surveys are inevitably contaminated by misidentified lower redshift sources.  The contamination could be particularly problematic for the VPF, which is exponentially sensitive to interlopers \citep{2008MNRAS.388.1101M}.  To gauge this effect, we selected a fraction of our LAEs and shuffled them to random locations in the field.  This procedure ignores the clustering of the interlopers, which is likely to be a small effect for the low number densities that we assume here. Based on spectroscopic followups, \citet{2018PASJ...70S..13O} estimated contamination rates of $\approx 8\%$ and $\approx 14\%$ in their $z=5.7$ and $z=6.6$ samples, respectively.   \citet{2018PASJ...70S..13O} also found that the contamination rate depends on source brightness.  They reported a higher rate of $\approx 33\%$ for sources brighter than 24 magnitudes in their $z=6.6$ sample. In Table \ref{tab:pval_table}, the p-values in parentheses assume contamination fractions of $10\%$ and $30\%$.  We find that the contamination significantly degrades the sensitivities of both the VPF and 2PCF.  However, for the 10-field configuration, we find that the VPF maintains an advantage over the 2PCF at $z=5.7$, even in the presence of signficant contamination.    

Our results suggest that the VPF can provide information on reionization that is complementary to the 2PCF.  We find that the former is a more sensitive test of the final stages of reionization in which neutral gas is contained primarily in the voids, and LAE surface densities are highest.

\subsection{Peak Statistics}
\label{sec:LAE_Groups}
LAE peak statistics offer a complementary picture to the voids.  When comparing models with $\bar{x}_{\rm HI} = 0$ and $0.3$ at fixed $\langle \Sigma \rangle$, any LAEs that are obscured in the voids of the latter model must be compensated by other LAEs unveiled elsewhere.  Thus, we also expect the statistics of LAE over-densities to be different between the two models (see \S \ref{sec:models}). We quantify this effect here.  We explore three different approaches, starting with the simplest statistic based on thresholding the LAE field.    

\subsubsection{Pixel Threshold Fraction}
\label{sec:ptf}

\begin{figure}
    \vspace{0cm}
    \includegraphics[width=8.5cm]{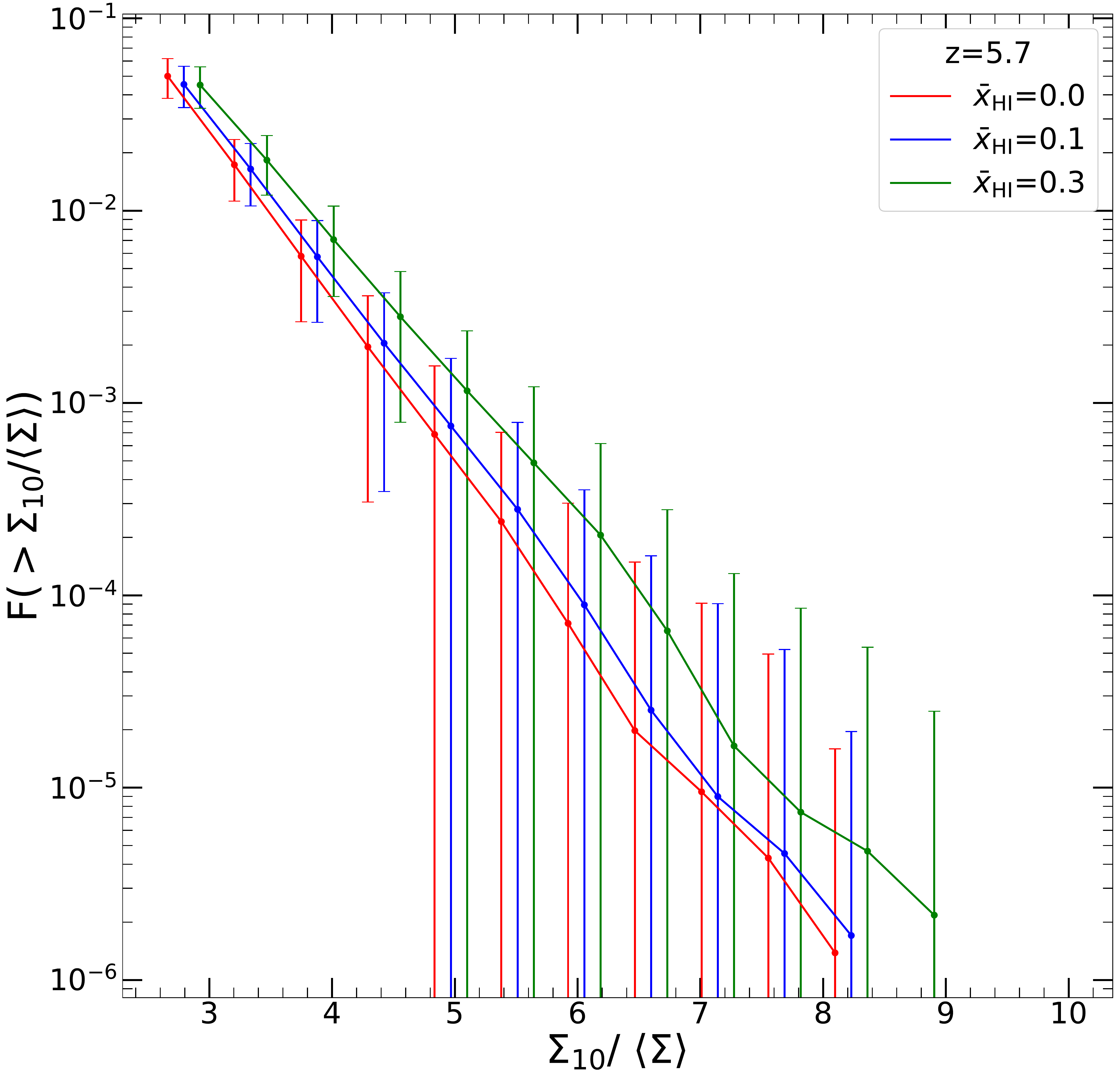}
    \vspace{0cm}
    \includegraphics[width=8.5cm]{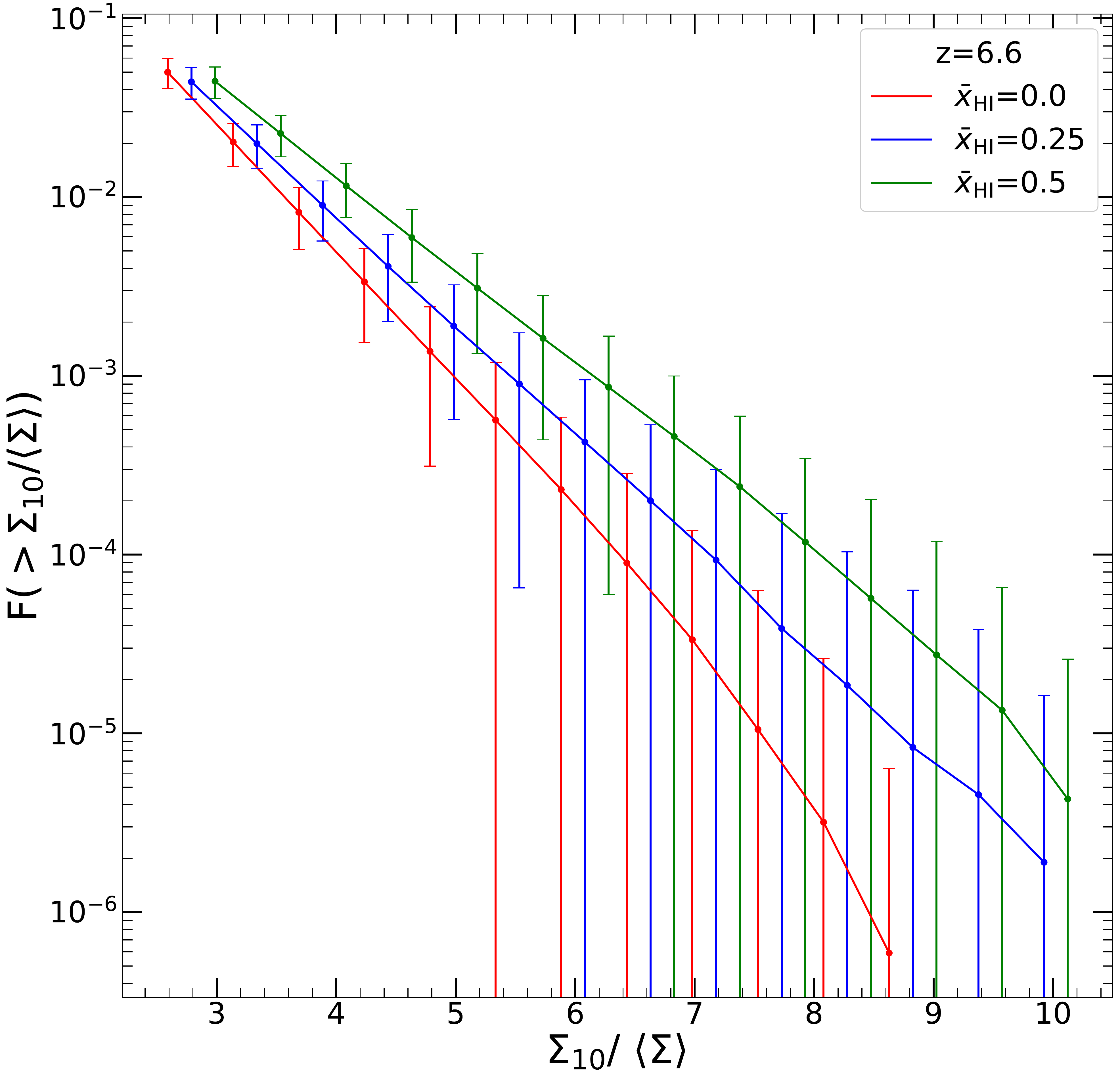}
    \caption{The Pixel Threshold Fraction (PTF), or fraction of pixels with LAE surface densities above $\Sigma_{10}/\langle \Sigma \rangle$.  Here, $\Sigma_{10}$ is the surface density smoothed over a scale of $\sigma= 10~h^{-1}$Mpc with a Gaussian filter. The error bars show the $1\sigma$ dispersion from cosmic variance in an expanded (10-field) SILVERRUSH-like survey.  At a given redshift, all models shown have a fixed $\langle \Sigma \rangle$ (see Table \ref{tab:beta_coeff}). In this case, a more neutral IGM leads to a larger fraction of over-dense pixels, a consequence of conservation of LAE number.  LAEs that are obscured in voids are compensated by more LAEs being visible elsewhere, which leads to an apparent enhancement in clustering.}
    \label{fig:pixel_stats}
\end{figure}

We begin by counting the fraction of pixels exceeding an over-density threshold in a smoothed map of the LAE distribution.  The LAE distribution is first smoothed onto a uniform 2-dimensional grid using a Gaussian filter with a smoothing scale of $\sigma = 10~h^{-1}$Mpc.  We employ a grid with $N=512^2$, but our conclusions are broadly robust to the exact choice. The size of the smoothing scale was adopted from previous studies of galaxy proto-clusters at $z=5.7$ and $6.6$ \citep[e.g.][]{2017ApJ...844L..23C}.  It is roughly the spatial extent of over-dense regions that later collapse into clusters at $z\sim 0$.   We then count the fraction of pixels above a given surface over-density $\Sigma_{10}(\mathbf{x})/\langle \Sigma \rangle$, where the subscript denotes that we have smoothed over a scale of $10~h^{-1}$Mpc.  We set the minimum threshold to be the 95th percentile in the $\bar{x}_{\rm HI} = 0$ models, since we find that differences between the models are strongest at the largest over-densities. For reference, this minimum threshold corresponds to 2$\langle \Sigma \rangle$ and 2.5$\langle \Sigma \rangle$ at $z=5.7$ and $6.6$, respectively.\footnote{Our thresholds are lower than those of previous studies which define clusters and proto-clusters with surface-densities greater than $4-5\sigma$ over the mean density \citep[e.g.][]{Inoue_2018, 2019ApJ...883..142H}.  In this section, we use the term ``LAE peak'' to distinguish between our liberal definition of a high-density region and the more stringent classification of proto-clusters used in those works \citep{Inoue_2018, 2019ApJ...883..142H}.}  We refer to this statistic, $F(> \Sigma_{10}/\langle \Sigma \rangle)$, as the Pixel Threshold Fraction (PTF). 

We show our results for the PTF in Figure \ref{fig:pixel_stats}. Consistent with expectations, the fraction of over-dense pixels increases with the neutral fraction for fixed $\langle \Sigma \rangle$.  This is particularly prominent amongst our models at $z=6.6$.   The differences in the PTF reflect an enhancement in the apparent clustering of LAEs due to patchy reionization, offering an alternative to the 2PCF for quantifying this effect. The error bars in Figure \ref{fig:pixel_stats} show the 1$\sigma$ cosmic variance obtained from the bootstrap method described in \S \ref{sec:voids}.  The results of our $\chi^2$ analysis on the PTF is also shown in Table \ref{tab:pval_table}.  At $z=6.6$, the PTF can discriminate the $\bar{x}_{\rm HI} = 0.5$ and $\bar{x}_{\rm HI} = 0.0$ models at a similar level to the 2PCF and the VPF. However, at $z=5.7$, the PTF is significantly weaker.  This behavior may owe to the fact that the LAEs in our models are more highly biased at $z=6.6$. 

\subsubsection{Peak Overdensity Function}
\label{sec:pof}

\begin{figure*}
    \centering
    \includegraphics[width=\textwidth]{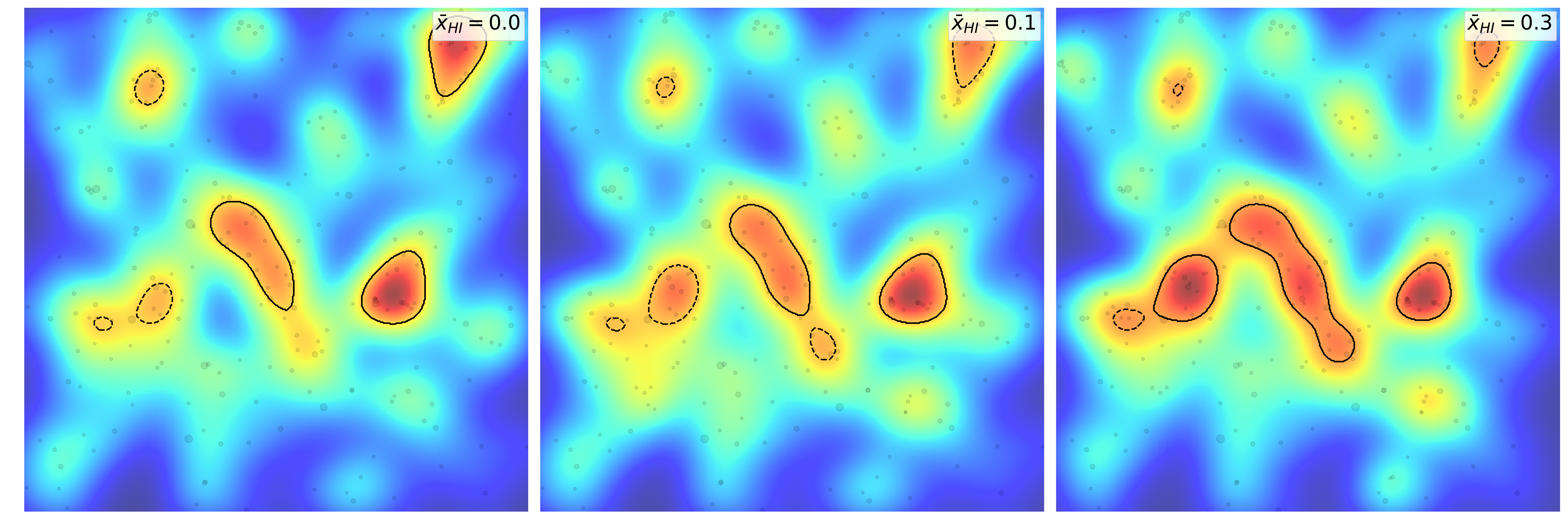}
    \includegraphics[width=\textwidth]{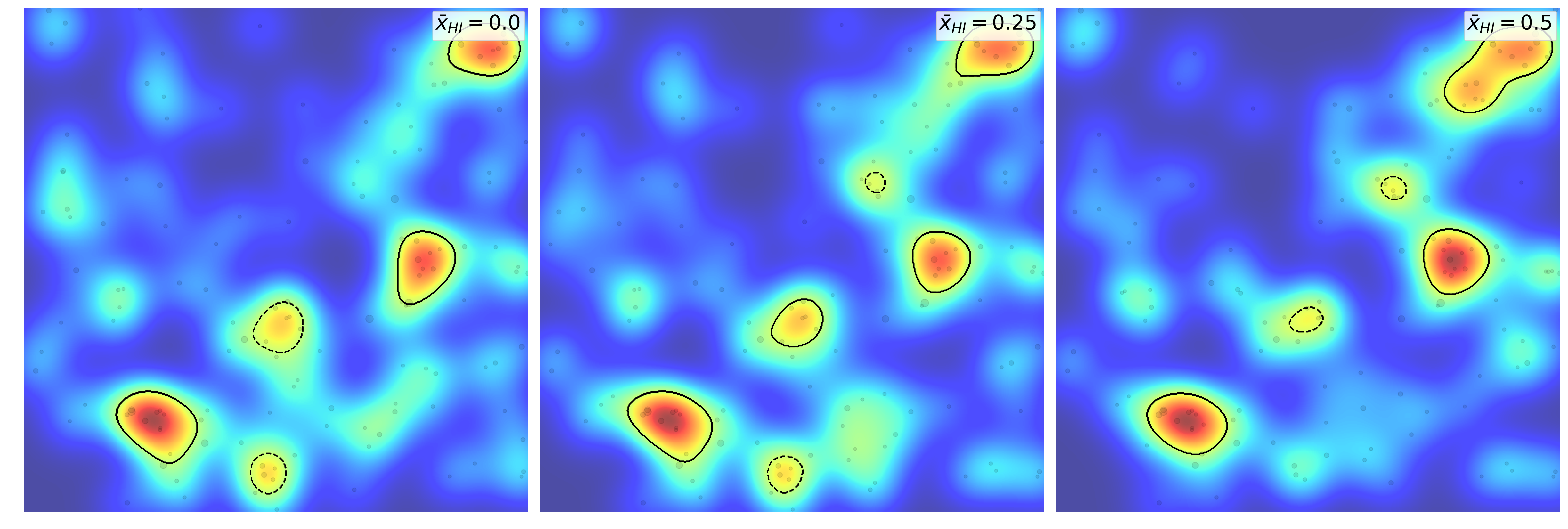} 
    \captionsetup{width=\textwidth}
    \caption{Examples of our peak finding algorithm described in \S \ref{sec:pof}. The color maps show the surface density contrast of the LAE field smoothed on a $\sigma = 10~h^{-1}$Mpc scale with a Gaussian filter.  The top and bottom rows correspond to $z=5.7$ and $z=6.6$, respectively, while the columns correspond to models with different global neutral fractions.  The top panels are 180$\times$180 $(h^{-1}\mathrm{Mpc})^2$ while the bottom panels are 200$\times$200 $(h^{-1}\mathrm{Mpc})^2$.  The solid black contours enclose regions identified as peaks in the LAE distribution. For reference, the dashed contours show over-dense regions that did not fulfill our LAE occupancy cut (see main text for details).}
    \label{fig:peak_map}
\end{figure*}

\begin{figure}
    \includegraphics[width=8.5cm]{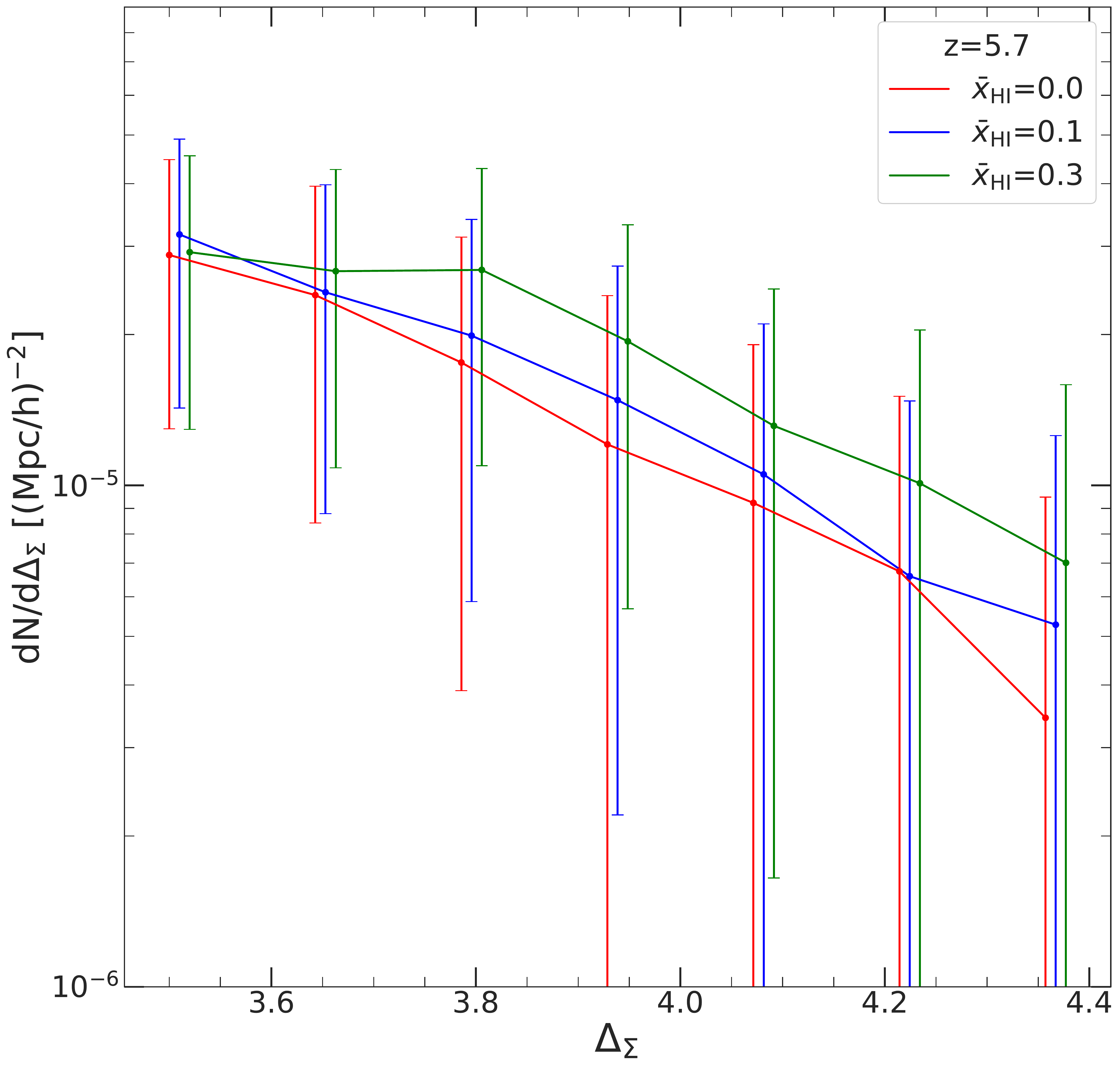}
    \includegraphics[width=8.5cm]{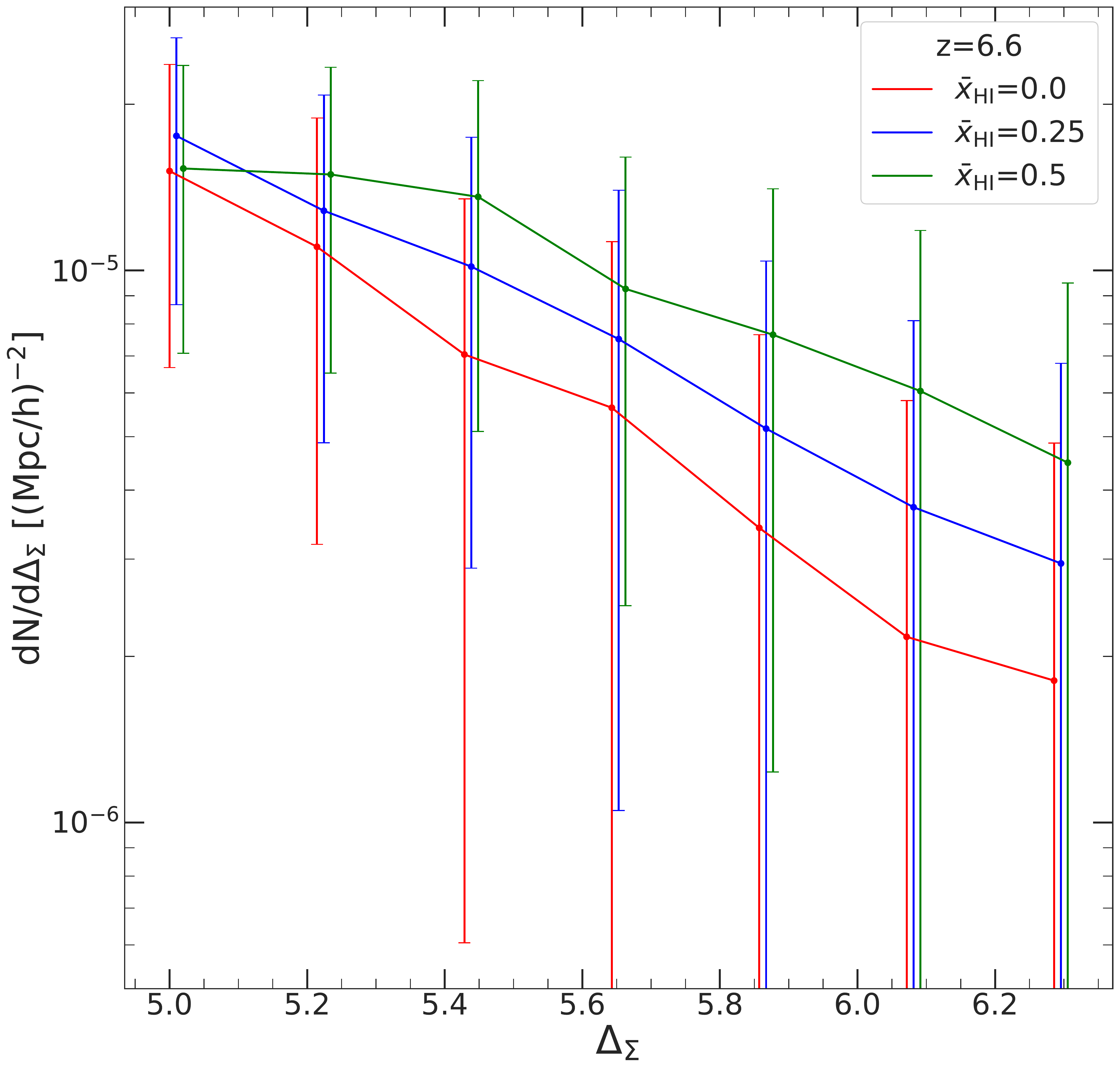}
    \caption{The Peak Overdensity Function (POF), $d\mathrm{N}/d\Delta_{\Sigma}$, or number of peaks per unit area, per unit surface over-density, $\Delta_{\Sigma} = \Sigma/\langle \Sigma\rangle$. For this statistic, peaks are identified as connected sets of pixels satisfying a minimum over-density threshold, as well as a minimum LAE occupancy (see main text for details).   Error bars show the $1\sigma$ dispersion from cosmic variance in an expanded (10-field) SILVERRUSH-like survey. At fixed $\langle \Sigma \rangle$, the over-dense LAE peaks are more numerous in models with higher neutral fraction.      }
    \label{fig:group_numbdens_stats}
\end{figure}

Next we consider a more sophisticated approach that groups together high surface density pixels (``peak pixels").  This approach has the added benefit that we can analyze the effect of reionization on the morphology of LAE peaks.   Peak pixels are identified in our smoothed surface density maps using the same threshold as in the last section. They are then grouped into connected sets that we call peaks.  To avoid edge effects, we eliminate peaks whose boundaries are one pixel from any edge of the field. We count the number of LAEs contained within each peak and eliminate peaks with less than $8(5)$ LAEs at $z=5.7(6.6)$, which, we find, reduces the effects of shot noise and grid resolution on our results.  This procedure yields catalogs of bounded regions of the highest intensity pixels. 
 
In Figure \ref{fig:peak_map}, we show example applications of our peak finding algorithm to mock fields at $z=5.7$ (top row) and $z=6.6$ (bottom row).  The columns correspond to different neutral fractions as denoted in the plot labels.  In each panel, the dots show the LAE distributions, while the color maps correspond to the smoothed surface density contrasts. The solid curves bound the LAE peaks selected by our algorithm while the dashed curves show regions that did not make our final cut owing to low LAE occupancy (i.e. not satisfying the minimum occupancy of 8(5) at $z=5.7$ and $6.6$). 

We have quantified the shapes of LAE peaks by computing moments of their surface density distributions.  We found that the shapes (e.g. ellipticities) and sizes of the peaks are statistically quite similar between our models, with variations $\sim 5 \%$. We do not consider these statistics further here.  Instead we focus on the abundance and occupancy of peaks, which do show significant differences. At fixed $\langle \Sigma \rangle$, a more neutral IGM leads to more LAEs per peak.  In particular, models with a higher neutral fraction contain more peaks with high LAE occupancy.  To demonstrate this, we consider the number of peaks per unit area, per unit surface density, $dN/d \Delta_{\Sigma}$, where $\Delta_{\Sigma}$ is the surface density of a peak in units of the global mean.   We term $dN/d \Delta_{\rm \Sigma}$ the Peak Overdensity Function (POF).  

Figure \ref{fig:group_numbdens_stats} shows the POF for our models at $z=5.7$ (top) and $z=6.6$ (bottom).  We find that $\Delta_{\Sigma}$ ranges from 3-5 at $z=5.7$ and 4-7 at $z=6.6$, generally reflecting the stronger bias of LAEs at higher redshifts.  The number of over-dense peaks clearly increases with the neutral fraction, consistent with our arguments above.  However, the error bars show considerable cosmic variance in even a future 10-field SILVERRUSH-like survey. Compared to the PTF, we find that the POF offers a much weaker discriminator between our models owing to the rarity of peaks. Indeed, Figure \ref{fig:peak_map} illustrates that only 2-3 peaks are identified per FoV.  Thus we did not consider the POF in our $\chi^2$ comparison.

\subsubsection{Peak Probability Function}
\label{sec:ppf}

\begin{figure}
    \includegraphics[width=8.5cm]{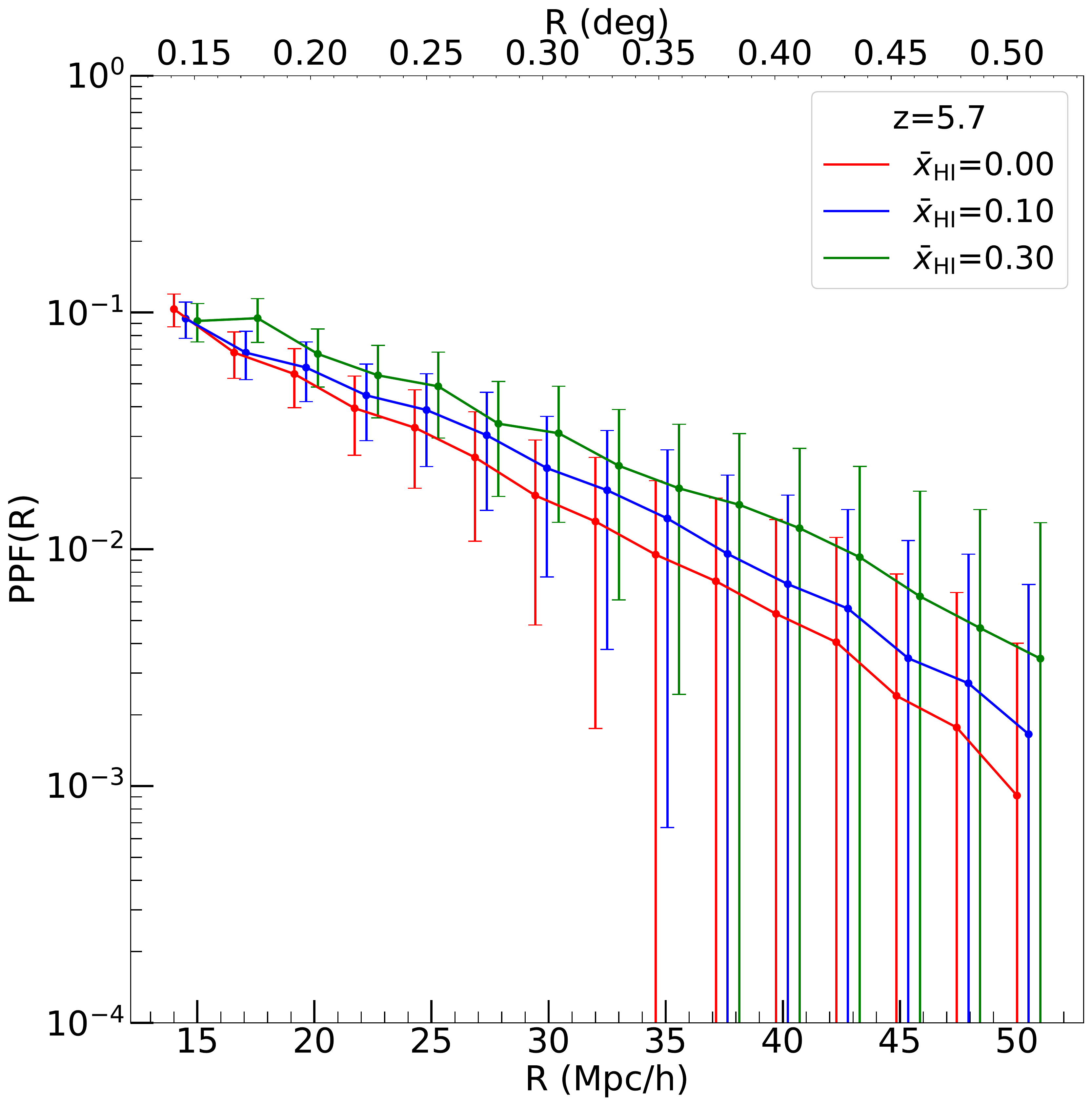}
    \includegraphics[width=8.5cm]{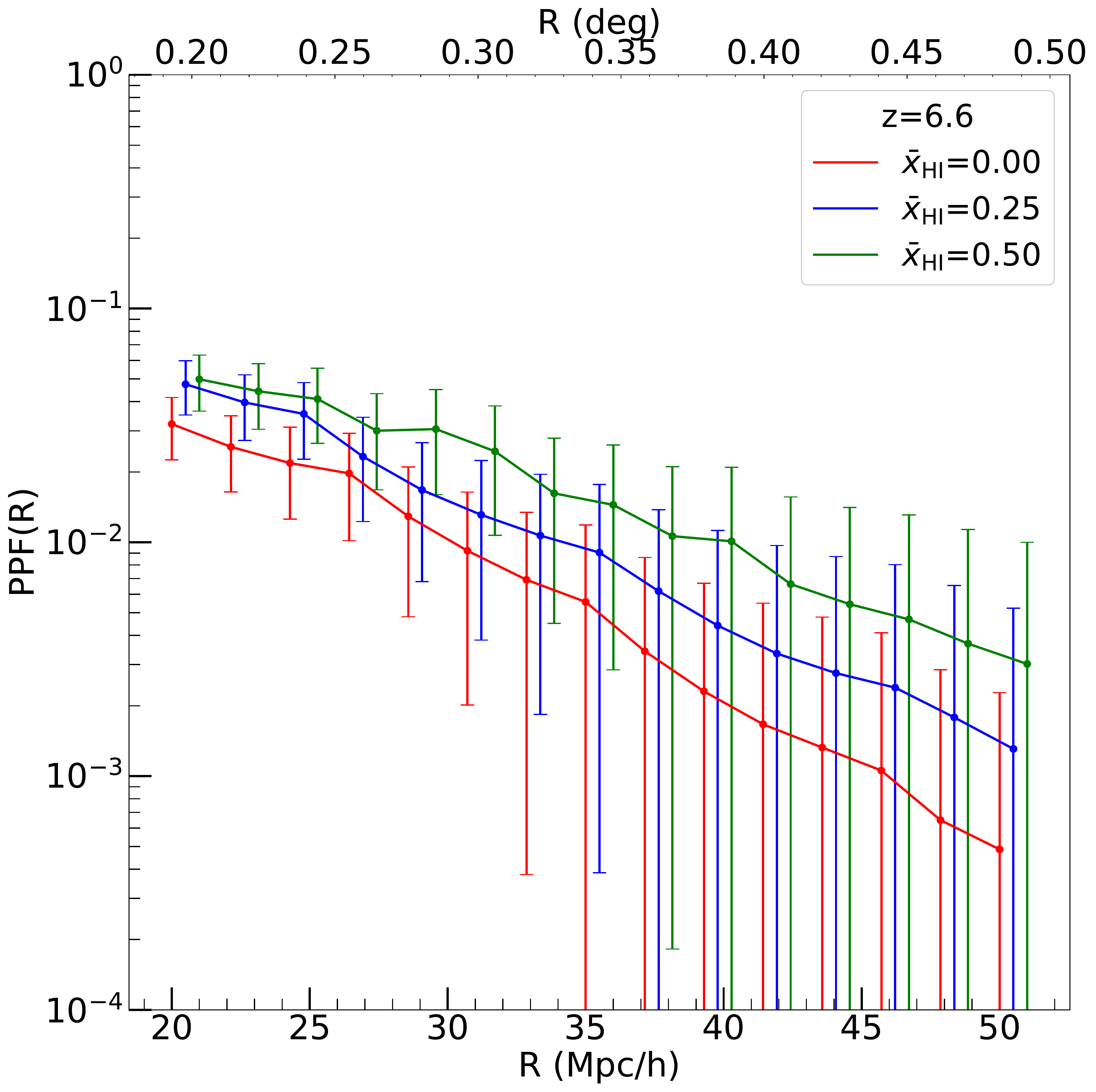}
    \caption{The Peak Probabiliy Function (PPF), or probability of finding an LAE over-density greater than some threshold within a circle of radius $R$.  We adopt a threshold of $2(2.5)\langle \Sigma \rangle$ at $z=5.7(6.6)$ (see main text for details). Error bars show the $1\sigma$ dispersion from cosmic variance in an expanded (10-field) SILVERRUSH-like survey. At fixed $\langle \Sigma \rangle$, higher neutral fractions yield increased probabilities of finding LAE peaks.}
    \label{fig:peak_prob_func}
\end{figure}

The last statistic that we consider is an analogue to the VPF of \S \ref{sec:voids}. We define the Peak Probability Function (PPF) to be the probability that a randomly placed circle of radius R has an over-density of LAEs greater than or equal to some threshold.  We randomly place down circles of radius R, find the number of LAEs ($N$) contained within the circle, and then calculate the surface density, $\Sigma_R = N/\pi R^2$. Using the same thresholds from \S \ref{sec:ptf}, we count only circles with $\Sigma_R > 2 (2.5) \langle \Sigma \rangle$ at $z=5.7(6.6)$.  We only consider R, such that there is a minimum of 10 LAE contained within a circle.  We then divide the number of circles making the cut by the total number drawn to find the probability as a function of $R$.

We show our results in Figure \ref{fig:peak_prob_func}.  Consistent with the previous statistics, we find that the probability of obtaining a peak increases with the neutral fraction.  This further cements the conclusion that, at fixed $\langle \Sigma \rangle$, a higher neutral fraction yields a higher abundance of over-dense peaks in the LAE distribution. The results of our $\chi^2$ analysis for the PPF are given in Table in Table \ref{tab:pval_table}.  Perhaps surprisingly, the PPF is not as strong of a test of our models as the other statistics we considered, and is significantly weaker than the 2PCF.  We attribute this finding to strong correlations between the radial bins, which are reflected in the highly off-diagonal structure of the covariance matrix for the PPF. The physical picture for this is as follows. Owing to the bias of LAEs, if a peak of a given radius $R$ is found, then the probability of finding additional peaks with other values of $R$ is enhanced significantly.  On other hand, the correlations between radial bins of the VPF are weaker. Finding a random hole in the LAE distribution does not necessarily enhance the probability of finding a larger hole, for example.

\section{Conclusion}
\label{sec:conc}
In this paper we have explored the use of narrow band LAE surveys to constrain late reionization scenarios, with an eye towards recently proposed models in which reionization ends as late as $z\approx 5$ \citep{2019MNRAS.485L..24K, 2019arXiv191003570N, 2020MNRAS.491.1736K}. During the tail end of reionization, with neutral fractions $\sim 10 \%$, the last remaining neutral islands in the Universe were likely relegated to voids in the galaxy distribution.  The neutral islands would have preferentially obscured LAEs in under-dense regions, motivating us to explore LAE void statistics further in this paper.  When comparing reionization models at a fixed mean surface density, LAEs obscured in the voids must be compensated by visible LAEs elsewhere in the spatial distribution.  Thus, we have also explored peak statistics as a complementary discriminator of reionization models. 

For several reionization scenarios spanning global neutral fractions of $\bar{x}_{\rm HI} = 0-0.3$ at $z=5.7$, and $\bar{x}_{\rm HI} = 0-0.5$ at $z=6.6$, we have constructed mock LAE surveys using one of the $L=(1~h^{-1}\mathrm{Gpc})^3$ Multi-Dark simulations as a basis. Our models were tuned to match the LAE surface densities observed in the recently published SILVERRUSH survey, as well as a number of other observational constraints.   The enormous simulation volume of Multi-Dark allowed us to capture the $\sim 10-100$ Mpc characteristic scales of reionization, and provided representative models for comparison against SILVERRUSH, which covers an area of $14-21$ deg$^2$ ($0.3-0.5$ Gpc$^2$).  One of our primary focuses was to quantify the cosmic variance expected in a SILVERRUSH-like survey. To this end we constructed a statistical ensemble of $1,000$ mock fields of view.    
 
 The SILVERRUSH data exhibits significant field-to-field variation in surface density, and some fields have visual asymmetries in the LAE distribution over $\sim 200$ Mpc scales.  We compared the SILVERRUSH data against our models and found that the surface density variations as well as the large-scale asymmetries of the former are well within expectations from large-scale structure formation alone. An exception to this conclusion is the field D-ELAISN1, which has $\approx 20 \%$ fewer LAEs than the lowest surface density observed in our ensemble.  Interestingly, D-ELAISN1 also exhibits among the largest half-plane asymmetries in its LAE distribution.  However, \citet{2018PASJ...70S..16K} has attributed the exceptional features of D-ELAISN1 to poor seeing, perhaps reconciling it with our models. We generally find that patchy reionization contributes only slightly to the cosmic variance and large-scale ($\sim 200$ Mpc) asymmetry of fields, and is an unlikely explanation for the characteristics of D-ELAISN1.         
 
 Next we considered the void probability function (VPF) as a probe of reionization models.  A more neutral IGM yields a higher probability of finding large-scale empty regions in the LAE distribution.  We found with a simple $\chi^2$ calculation that the VPF is a more sensitive test of models during the last stages of reionization than the widely used angular two-point correlation function (2PCF). In spite of the VPF being exponentially sensitive to interlopers, this conclusion holds even if we assume that $10-30 \%$ of the sources are misidentified foreground objects.  
 
 We also explored three approaches to quantifying the statistics of peaks in the LAE distribution. Using a method that groups over-dense pixels in surface density maps smoothed over $10~h^{-1}$Mpc scales, we found that patchy reionization has little effect statistically on the shapes of LAE peaks. When comparing models at fixed mean surface density, the predominant effect of neutral gas is to increase the probability of finding LAE over-densities.  We found that this effect can be quantified straightforwardly by thresholding smoothed LAE maps and counting the fraction of over-dense pixels. Our simplistic $\chi^2$ analysis showed that this peak statistic offers roughly the same sensitivity as the 2PCF and VPF for discriminating between models with $\bar{x}_{\rm HI} = 0.0$ and $\bar{x}_{\rm HI} = 0.5$.  However, it is generally less sensitive during the last stages of reionization. 
 
 In this work, we have focused on constraining the latter half of reionization with a SILVERRUSH-like survey. However, SILVERRUSH and The Lyman Alpha Galaxies in the Epoch of Reionization (LAGER; \citealt{2017ApJ...842L..22Z}) project will continue to expand the sample of known LAEs at $\geq 7$.  Future work should address in further detail the efficacy for $z\sim 7$ LAEs  to constrain reionization, as well as the trade-off between survey area and depth (see Perez et al. in prep).   We note that all of our calculations have assumed that reionization proceeded in an inside-out manner, consistent with semi-numeric models and more detailed radiative transfer simulations. Future studies might further explore whether a combination of statistics could provide more fundamental insight into the connection between reionization and the underlying density field.  
 
 In recent years, the sample of known LAEs at $z \geq 5.7$ has grown to over $2,000$, raising the prospects for detecting the last stages of reionization in the clustering of LAEs.  Our calculations suggest that the statistics of voids and peaks offer complimentary approaches to the 2PCF, and may even be more optimal under some conditions.

\section*{Acknowledgements}

We thank Lucia Perez, George Becker, Simeon Bird, and Matt McQuinn for useful discussions and/or comments on this manuscript. We are especially grateful to Hy Trac for providing his RadHydro radiation hydrodynamics code. A.D. acknowledges support from HST award HST-AR15013.005-A and NASA award 19-ATP19-0191.  Z.Z. is supported by NSF grant AST-2007499.  This work made use of the following software packages: \verb|matplotlib| \citep{Hunter:2007}, \verb|numpy| \citep{harris2020array}, \verb|pandas| \citep{mckinney-proc-scipy-2010}, \verb|scipy| \citep{2020SciPy-NMeth}, and \verb|scikit-image| \citep{scikit-image}.  We additionally made use of the software package \verb|CosmoloPy|, which can be found at this URL \url{http://roban.github.com/CosmoloPy/}.  The authors recognize the cultural and religious importance of the Mauna Kea summit to the indigenous Hawaiian community.




\bibliographystyle{mnras}
\bibliography{master} 

\appendix
Here is a Table providing the split screen variances we found for all the SILVERRUSH fields:




\bsp	
\label{lastpage}
\end{document}